\documentclass[twocolumn,showpacs,preprintnumbers,amsmath,amssymb]{revtex4}
\usepackage{graphicx,epsfig}
\usepackage{dcolumn}
\usepackage{bm}
\newcommand{\bi}{\begin{itemize}}
\newcommand{\ei}{\end{itemize}}
\newcommand{\be}{\begin{equation}}
\newcommand{\ee}{\end{equation}}
\newcommand{\ba}{\begin{eqnarray}}
\newcommand{\ea}{\end{eqnarray}}
\newcommand{\bse}{\begin{subequations}}
\newcommand{\ese}{\end{subequations}}
\newcommand{\ACal}{{\cal{A}}}

\newcommand{\DD}{{\cal {D}}}

\newcommand{\bben}{\begin{itemize}}
\newcommand{\eeen}{\end{itemize}}
\newcommand{\bbq}{\begin{quote}}
\newcommand{\eeq}{\end{quote}}

\newcommand{\RR}{{}^3{\cal{R}}}

\newcommand{\T}{{}^3{\cal{T}}}

\newcommand{\EE}{{\cal{E}}}

\newcommand{\JJ}{{\cal{J}}}

\newcommand{\HH}{{\cal{H}}}
\newcommand{\KK}{{\cal{K}}}
\newcommand{\PP}{{\cal{P}}}
\newcommand{\MM}{{\cal{M}}}

\newcommand{\Dih}{\delta_i^{(\HH)}}
\newcommand{\Dim}{\delta_i^{(\mu)}}
\newcommand{\Dik}{\delta_i^{(\KK)}}
\newcommand{\Diq}{\delta_i^{(q)}}

\newcommand{\Da}{\delta^{(A)}}

\newcommand{\Dh}{\delta^{(\HH)}}
\newcommand{\dDh}{\dot\delta^{(\HH)}}
\newcommand{\Dth}{\delta^{(\Theta)}}

\newcommand{\Dm}{\delta^{(\mu)}}
\newcommand{\dDm}{\dot\delta^{(\mu)}}
\newcommand{\Dk}{\delta^{(\KK)}}
\newcommand{\Dp}{\delta^{(p)}}
\newcommand{\Dq}{\delta^{(q)}}
\newcommand{\dDq}{\dot\delta^{(q)}}
\newcommand{\Drho}{\delta^{(\rho)}}
\newcommand{\dDrho}{\dot\delta^{(\rho)}}

\newcommand{\dd}{{\rm{d}}}

\begin{document}


\title{Quasi--local variables in spherical symmetry: numerical applications to dark matter and dark energy sources.} 
\author{ Roberto A. Sussman$^\ddagger$}
\affiliation{
$^\ddagger$Instituto de Ciencias Nucleares, Universidad Nacional Aut\'onoma de M\'exico (ICN-UNAM),
A. P. 70--543, 04510 M\'exico D. F., M\'exico. }
\date{\today}
\begin{abstract} A numerical approach is considered for spherically symmetric spacetimes that generalize Lema\^\i tre--Tolman--Bondi dust solutions to nonzero pressure (``LTB spacetimes''). We introduce quasi--local (QL) variables that are covariant LTB objects satisfying evolution equations of Friedman--Lema\^\i tre--Robertson--Walker (FLRW) cosmologies. We prove rigorously that relative deviations of the local covariant scalars from the QL scalars are non--linear, gauge invariant and covariant perturbations on a FLRW formal ``background'' given by the QL scalars. The dynamics of LTB spacetimes is completely determined by the QL scalars and these exact perturbations. Since LTB spacetimes are compatible with a wide variety of ``equations of state'', either single fluids or mixtures, a large number of known solutions with dark matter and dark energy sources in a FLRW framework (or with linear perturbations) can be readily examined under idealized but non--trivial inhomogeneous conditions.  Coordinate choices and initial conditions are derived for a numerical treatment of the perturbation equations, allowing us to study non--linear effects in a variety of phenomena, such as gravitational collapse, non--local effects, void formation, dark matter and dark energy couplings and particle creation.  In particular, the embedding of inhomogeneous regions can be performed by a smooth matching with a suitable FLRW solution, thus generalizing the Newtonian ``top hat'' models that are widely used in astrophysical literature. As examples of the application of the formalism,  we examine numerically the formation of a black hole in an expanding Chaplygin gas FLRW universe, as well as the evolution of density clumps and voids in an interactive mixture of cold dark matter and dark energy.           
\end{abstract}

\maketitle
\section{Introduction.}

Recent observations suggest that present cosmic dynamics is dominated by elusive sources called ``dark matter'' and ``dark energy'', the former clustered in galactic halos and the latter possibly associated with a large scale repulsive (yet unknown) interaction. A large body of phenomenological and theoretical models have been proposed to describe these sources (see \cite{review} for a comprehensive review). The dominant model of dark matter is a collision--less gas of supersymmetric particles: ``cold'' dark matter~\cite{review}, whereas dark energy has been described by a  cosmological constant \cite{Lambda}, as well as by quintessence scalar fields, Phantom fields, tachyons, branes (see \cite{review}). A unified perspective is given by the Chaplygin gas~\cite{review,chaplygin1,chaplygin2}, a single source that exhibits the expected dynamical behavior of dark matter and dark energy. There is also a large number of empiric and phenomenological descriptions of dark matter interacting with dark energy \cite{intmix1,intmix2} (see \cite{intmix_new} for a recent appraisal), as well as empiric dark energy ``equations of state'' that fit observations~\cite{EOS1,EOS2}.  

While dark matter at the galactic scale is considered inhomogeneous, due to its association with structure formation, most research work on dark energy sources has been conducted in homogeneous Friedman--Lema\^\i tre--Robertson--Walker (FLRW) spacetimes and/or linear perturbations, since dark energy is assumed to have a significant effect only at a larger cosmic scale in which the universe appears to be homogeneous.  However, the possibility of anisotropic or inhomogeneous dark energy (or combined dark matter and energy sources) is beginning to be discussed in the literature \cite{chapinhom,chapinhom2,chapstar,intmix3,Mota_anis}.  Given our ignorance on the fundamental properties of these sources, there is no theoretically binding reason to assume, {\it a priori}, that no valuable new information would result from studying their interaction under inhomogeneous conditions, at least in scales associated with structure formation and gravitational clustering.

Alternative proposals to the dark energy paradigm also consider a full relativistic treatment of cosmological inhomogeneities~\cite{Inh_rew1,Inh_rew2}. Numerous articles describe the possibility that cosmic acceleration (or at least part of it) could result from the presence of inhomogeneites in photon trajectories in observations from high red--shift objects~\cite{InhObs1, InhObs2}. Following a more theoretical perspective, various averaging formalisms~\cite{ave1,ave2} have examined the occurrence of an effective acceleration from the so--called ``back--reaction'', associated with non--local effects that emerge as inhomogeneous sources are averaged (see \cite{theo1,theo2} for further discussion and \cite{ltbave,condsBR} for scalar averages in the spherical dust case). Also, an interesting  connection  between cosmic acceleration, back--reaction and non--trivial gradients of quasi--local energy has been proposed~\cite{wiltshire}, which could clarify important theoretical  issues (the connection between the material presented in this paper and back--reaction issues is discussed in \cite{condsBR,QLvars}). Possible observational consequences of these theoretical alternatives to dark energy are certainly worth serious consideration~\cite{Obs_BR} (see also \cite{InhObs1,InhObs2} ).  Even if dark energy prevails over these alternative proposals, there is no harm done (and perhaps valuable new information) in probing its behavior under inhomogeneous conditions.

Ideally, the study of inhomogeneous sources should be conducted on  spacetimes not restricted by simplifying symmetries, but this would require fully 3--dimensional codes of high complexity. As a compromise between linear perturbations and this type of ``realistic'' generality, we present in this article a class of inhomogeneous spherically symmetric models that can be well handled with relatively simple numerical methods. Although these models are idealized, they are non--trivial, exhibit non--linear behavior, and are general enough for testing a wide variety of physical assumptions on modern cosmological sources. 

By assuming a spherically symmetric Lema\^\i tre--Tolman--Bondi metric, we derive in section \ref{genLTB} a class of spacetimes (``LTB spacetimes'') that generalize to nonzero pressure (and nonzero pressure gradients) the well know solutions for inhomogeneous dust associated with this metric \cite{kras}. The most general source for this metric (in the comoving frame) is a fluid with anisotropic stresses, which will end up being interpreted as fluctuations of the isotropic pressure. However, regardless of the interpretation, we remark that the existence of pressure anisotropy is far from a drawback or shortcoming, not only because it can be associated to a number of well motivated physical effects~\cite{anisP}, but because an inhomogenous source with only isotropic pressure is a far more idealized situation than one with anisotropic pressure.   

In section \ref{1plus3} we show that LTB spacetimes can be fully characterized by a set of local covariant scalars, hence their dynamics becomes completely determined by solving the evolution equations for these scalars in the covariant fluid flow (or ``1+3'') representation~\cite{ellisbruni89,BDE,LRS,1plus3}. By looking at the relation between energy density and the Misner--Sharp quasi--local mass--energy invariant~\cite{MSQLM,kodama,szab,hayward1,hayward2}, we introduce in section \ref{QLdefs} ``quasi--local'' (QL)  scalar functions, which are LTB objects that satisfy FLRW dynamics. Expressing all local covariant scalars as perturbations of the QL scalars, we rewrite in section \ref{eveqs} the 1+3 fluid flow equations of section \ref{1plus3} as evolution equations for the QL scalars and these perturbations, which on the basis of known criteria that define a perturbation formalism~\cite{ellisbruni89, BDE,LRS,1plus3,bardeen}, are shown in section \ref{perturb} to be covariant, gauge--invariant, non--linear perturbations on a FLRW formal ``background'' given by the QL variables. These evolution equations are equivalent to ODE's in which radial dependency enters as a parameter (see Appendix C of \cite{suss08})).

As shown in section \ref{EOS}, once we choose an ``equation of state'' (EOS) between the QL energy density and pressure (which are ``background'' variables), the QL evolution equations become determined. We point out that, by choosing such an EOS, correlation terms appear in the relations between local scalars, in a similar way as virial corrections to the ideal gas EOS appear in Newtonian self--gravitating systems~\cite{saslaw1,saslaw2}. These terms can be associated with the long range nature of gravity~\cite{Padma_GTD}. Since the fundamental physics of dark matter and dark energy sources is still unknown, we cannot rule out the possibility that this type of non--local effects could play an important dyamical role when these sources are studied under non--trivial and non--linear inhomogeneity. 

In section \ref{mixtures} we apply the perturbation formalism to a fluid mixture, which can be interactive or with each component separately conserved (decoupled). This mixture description can be readily used to generalize to inhomogeneous conditions similar mixture models derived in a FRLW context in the many references quoted in \cite{intmix1,intmix2,intmix_new}. We show in this section that scalar fields can only be compatible with LTB spacetimes in mixture sources, playing the role of the homogeneous dark energy component (coupled with inhomogeneous dark matter~\cite{intmix3}) 

Coordinate choices appropriate for the numerical treatment for the evolution equations of sections \ref{eveqs} and \ref{mixtures} are discussed in \ref{FLRWlike}, while in section \ref{FLRWsubs} we show that perturbation functions are well defined as long as shell crossing singularities are absent. A possible (but not compulsory) way in which LTB spacetimes can be embedded into a FLRW background is by smoothly matching an inhomogeneous LTB section with a section of a suitable FLRW spacetime. We provide in section \ref{tophats} the conditions for such a matching, leading to smooth and fully relativistic generalizations of the Newtonian ``top hat'' models that are widely used in the astrophysical literature~\cite{tophats}. 

We examine in section \ref{applications} two examples of the application of the formalism presented in the previous sections. In the first example, we solve numerically the evolution equations for a smooth and fully relativistic Chaplygin gas ``top hat'' model, in which a black hole is formed in an inhomogeneous comoving section of the LTB  Chaplygin gas, smoothly embedded in an expanding Chaplygin gas FLRW universe. In the second example, we examine a mixture of interactive dark matter and dark energy of the type examined in \cite{intmix1,intmix2,intmix_new}, but give the interaction a phenomenological interpretation in terms of particle creation, with dark matter particles decaying into dark energy~\cite{intmix_new,lima}. We show how the radial profiles of dark matter density evolve from initial clumps into deep voids, thus providing a toy model for a void formation scenario in the context of this type of interactions. In the appendices we provide detailed information on how initial conditions and coordinate choices can be set up for the numerical integration of the evolution equations.

\section{Generalized LTB spacetimes}\label{genLTB}

We will denote by ``LTB spacetimes'' all spherically symmetric spacetimes that are solutions of Einstein's equations  described by the Lemaitre--Tolman--Bondi metric (in the comoving frame):
\begin{equation} ds^2 = -c^2dt^2 +\frac{R'^2\,dr^2}{1-K}+R^2[d\theta^2+\sin^2\theta\,d\phi^2],\label{LTB}\end{equation}
where $R=R(ct,r)$, \, $R'=\partial R/\partial r$ and $K=K(r)$. The metric (\ref{LTB}) is normally associated with a dust source, but the most general source compatible with LTB spacetimes is an anisotropic fluid~\cite{kras}
\begin{equation}T^{ab} = \mu\,u^a u^b + p\,h^{ab}+\Pi^{ab},\label{Tab}\end{equation}  
where $u^a=\delta^a_0$,\,with $\mu(ct,r),\,p(ct,r)$ being respectively the energy density and isotropic pressure, $\Pi^{ab}$ is the anisotropic pressure tensor and $h^{ab}=u^au^b+g^{ab}$ is the induced metric of the hypersurfaces $\T(t)$ orthogonal to $u^a$, marked by constant $t$.

The covariant objects associated with LTB spacetimes are the scalars $\mu,\,p$, plus the expansion and Ricci scalar of the $\T(t)$ (respectively $\Theta$ and $\RR$)
\ba \mu &=& u_au_bT^{ab},\qquad p =\frac{1}{3}h_{ab} T^{ab},\\
 \Theta &=& \tilde\nabla_a u^a = \frac{2\dot R}{R}+\frac{\dot R'}{R'},\label{theta1}\\
\RR  &=& \frac{2(KR)'}{R^2R'},\label{RR1}\ea
together with the anisotropic pressure $\Pi^{ab}$, shear $\sigma^{ab}$ and $E^{ab}$ Electric Weyl tensors
\ba \Pi^{ab} &=& \left[h^{(a}_ch^{b)}_d-\frac{1}{3}h^{ab}h_{cd}\right]\, T^{cd},\label{PPtens}\\
\sigma_{ab} &=& \tilde\nabla_{(a}u_{a)}-\frac{\Theta}{3}h_{ab},\label{sigma1}\\
 E^{ab} &=& u_c u_d C^{abcd},\label{EEtens}\ea
where $h_c^{(a} h_d^{b)}$ denotes symmetrization on $a,b$,\, $\dot R=u^a\nabla_a R=\partial R/\partial ct$,\, $\tilde\nabla_a=h_a^b\nabla_b$ and $C^{abcd}$ is the Weyl curvature tensor.

Each spacelike symmetric trace--less tensors like $\Pi^{ab},\,\sigma^{ab},\,E^{ab}$ in LTB spacetimes can be expressed in covariant manner in terms of a single scalar function as 
\bse\label{PSEsc}\ba \Pi^{ab} &=& \PP\,\Xi^{ab},\qquad\Rightarrow\quad \PP = \Xi_{ab}\Pi^{ab},\label{PPsc}\\ 
\sigma^{ab}&=&\Sigma\,\Xi^{ab},\qquad\Rightarrow\quad \Sigma = \Xi_{ab}\sigma^{ab},\label{Sigsc}\\ E^{ab}&=&
\EE\,\Xi^{ab},\qquad\Rightarrow\quad \EE = \Xi_{ab}E^{ab},\label{EEsc}\ea\ese
where the tensor  $\Xi^{ab}\equiv h^{ab}-3\chi^{a}\chi^b$ and $\chi^a=\sqrt{h^{rr}}\,\delta^a_r$ is the unit vector orthogonal to $u^a$ and to the orbits of SO(3) parametrized by $(\theta,\phi)$. 

The scalars $\mu,\,p,\,\PP$ relate to the metric functions by means of the field equations $G^{ab}=\kappa T^{ab}$ (with $\kappa =8\pi G/c^4$) for (\ref{LTB})--(\ref{Tab}):
\bse\label{fieldeqs}\ba \kappa\,\mu\,R^2R' &=& \left[R(\dot R^2+K)\right]',\label{mu1}\\
\kappa\,p\,R^2R' &=& -\frac{1}{3}\left[R(\dot R^2+K)+2R^2\ddot R\right]',\label{p1}\\
\kappa\,\PP\,\frac{R'}{R} &=& -\frac{1}{6}\left[\frac{\dot R^2+K}{R^2} +\frac{2\ddot
Y}{Y}\right]',\label{PP1}\ea\ese
while from (\ref{theta1}) and (\ref{PSEsc}), we obtain for $\EE$ and $\Sigma$ 
\bse\label{ESsc}\ba\Sigma &=& \frac{1}{3}\left[\frac{\dot R}{R}-\frac{\dot R'}{R'}\right],\label{Sigma1}\\
\EE &=& -\frac{\kappa}{2}\,\PP-\frac{\kappa}{6}\,\mu+ \frac{\dot
R^2+K}{2R^2}.\label{EE1}\ea\ese
It is evident that LTB spacetimes are completely characterized by the following set of covariant scalars (see \cite{LRS} for comparison):
\begin{equation} \{\mu,\,p,\,\PP,\,\Theta,\,\Sigma,\,\EE,\,\RR\},\label{loc_scals}\end{equation}
which we will denote as the ``local'' scalar representation. 

\section{Field equations vs. fluid flow equations}\label{1plus3}

In principle, LTB specetimes can be determined by solving the field equations after supplying an ``equation of state'' (EOS) linking $\mu,\,p$ and $\PP$. This can be illustrated by looking at the best known LTB spacetime, a dust source with the simple EOS: $p=\PP=0$, for which total energy density is rest--mass density: $\mu=\rho c^2$. In this case,  (\ref{mu1})--(\ref{PP1}) reduce to
\ba \kappa \rho c^2 R^2R' = 2M',\label{rho_dust}
\\ \dot R^2 = \frac{2M}{R}-K,\label{fried_dust}
\ea
where $M=M(r)$ appears as an ``integration constant'' and the rest of the scalars $\Theta,\,\Sigma,\, \RR,\, \EE$ remain the same (with $\PP=0$ in (\ref{EE1})). The standard tactic in this case is to solve the Friedman equation (\ref{fried_dust}) for a given choice of $M(r)$ and $K(r)$, and then to evaluate from this solution all the derivatives of $R,\,M,\,K$ in order to compute the scalars $\rho,\,\Theta,\,\Sigma,\, \RR,\, \EE$  from (\ref{theta1}), (\ref{RR1}), (\ref{Sigma1}), (\ref{EE1}) and (\ref{rho_dust}). There is an extensive literature on how this is done (see \cite{kras,ltbstuff,suss02}), and the same tactic can work on simple particular cases with non--zero pressure (see \cite{kras,hydro}) and simple EOS relating $\mu,\,p$ and $\PP$, but for a general LTB spacetime it is not possible (or very difficult) to find and solve an equation like (\ref{fried_dust}) in order to evaluate radial gradients of $R$ for computing the scalars in (\ref{loc_scals}). 

An alternative approach to deal with LTB spacetimes follows from the evolution equations (or ``fluid flow'' equations), based on the covariant ``1+3'' decomposition of Ehlers, Ellis, Bruni, Dunsbury and Van Elst~\cite{ellisbruni89,BDE,LRS,1plus3}. Instead of working on the relation between (\ref{loc_scals}) and the metric functions, in this approach the field and conservation equations $\nabla_a T^{ab}=0$ for (\ref{LTB})--(\ref{Tab}) are transformed into an equivalent first order system of time evolution equations and constraints for (\ref{loc_scals}). In particular, the spherically symmetric case is discussed in \cite{LRS}. 

Since the expansion and 3--dimensional Ricci scalars very often appear as $\Theta/3$ and $\RR/6$, the notation becomes greatly simplified if we define:
\begin{equation}\HH \equiv \frac{\Theta}{3},\qquad \KK \equiv \frac{\RR}{6},\label{HKdefs}\end{equation}
Bearing in mind (\ref{PPsc})--(\ref{EEsc}) and (\ref{HKdefs}), the ``1+3'' equations for LTB spacetimes become the following set of scalar evolution equations
\bse\label{eveqs_13}\ba
\dot\HH &=&-\HH^2
-\frac{\kappa}{6}\left(\,\mu+3p\,\right)-2\Sigma^2,\label{ev_theta_13}\\
\dot \mu &=& -3\,(\mu+p)\,\HH-6\,\Sigma\, \PP,\label{ev_mu_13}\\
\dot\Sigma &=& -2\,\HH\,\Sigma+\Sigma^2-\EE+\frac{\kappa}{2}\PP,
\label{ev_Sigma_13}\\
 \dot\EE &=& -\frac{\kappa}{2}\dot \PP-\frac{\kappa}{2}\left[\mu+p-2\PP\right]\Sigma
-3\left(\EE+\frac{\kappa}{6}\PP\right)(\HH+\Sigma),\nonumber\\\label{ev_EE_13}\ea\ese
together with the spacelike constraints  
\bse\label{c_13}\ba (p-2\PP)\,'-6\,\PP\,\frac{R'}{R}=0,\label{cPP_13}\\
\left(\Sigma+\HH\right)'+3\,\Sigma\,\frac{R'}{R}=0,\label{cSigma_13}\\
\frac{\kappa}{6}\left(\mu+\frac{3}{2}\PP\right)'
+\EE\,'+3\,\EE\,\frac{R'}{R}=0,\label{cEE_13}\ea\ese
and the Friedman equation (or ``Hamiltonian'' constraint)
\begin{equation}\HH^2 = \frac{\kappa}{3}\, \mu
-\KK+\Sigma^2,\label{cHam_13}\end{equation}
The system
(\ref{ev_theta_13})--(\ref{cHam_13}) still requires an equation of state (EOS) linking $\mu,\,p$ and
$\PP$ to become determined. However, time and radial derivatives do not decouple, in general.
Moreover, (\ref{loc_scals}) is not the only scalar representation for LTB spacetimes. 

\section{The quasi--local (QL) scalars.}\label{QLdefs} 

The Misner--Sharp QL mass--energy function, $\MM$, is a well known invariant defined for spherically symmetric spacetimes as \cite{MSQLM,szab,kodama,hayward1,hayward2}:
\ba \frac{2\MM}{R} \equiv R^2{\cal{R}}^{\theta\phi}\,_{\theta\phi} =
1-\frac{(\partial_r R)^2}{g_{rr}}-\frac{(\partial_0 R)^2}{g_{00}},\label{MSQLM}\ea
where ${\cal{R}}^{ab}\,_{cd}$ is the Riemann tensor. For LTB spacetimes with an anisotropic tensor like (\ref{Tab}), $\MM$ defined by (\ref{MSQLM}) satisfies:
\bse\label{QLM_expr}\ba 
 2\MM' &=& \kappa \mu R^2 R', \label{QLM1}\\
 2\dot\MM &=& -\kappa (p-2\PP) R^2 \dot R,\label{QLM2} \ea\ese
whose integrability condition is equivalent to the conservation law $\nabla_b T^{ab}=0$ (equations (\ref{ev_mu_13}) and (\ref{cPP_13})). In fact, the scalars $\MM$ and $R$ are the main invariants in spherically symmetric sacetimes \cite{hayward1,hayward2}, hence all objects constructed with $R$ and $\MM$ and their covariant derivatives are necessarily covariant objects (see \cite{szab,kodama}).  

Comparing (\ref{mu1}) with (\ref{QLM1}) suggests that $\MM$ can be equal to the integral of $\mu$ in  (\ref{mu1}) as long as this integral can be performed and is finite.  If we assume the existence of a regular symmetry center at $r=0$, then we can integrate both sides of (\ref{mu1}) along comoving domains of the hypersurfaces $\T(t)$ to obtain
\begin{equation} \frac{2\MM}{R^3}=\frac{\kappa}{3}\,\mu_*=\frac{\kappa}{3}\frac{\int_{x=0}^{x=r}{\mu\,R^2R' dx}}{\int_{x=0}^{x=r}{R^2R' dx}}=\frac{\dot R^2+K}{R^2},\label{QLM_mu}\end{equation}
which, for spherical symmetry, characterizes $\MM$ as the ``effective'' mass--energy in a comoving volume~\cite{kodama,hayward1,hayward2} and the ``ADM'' mass in the appropriate asymptotic limits~\cite{MSQLM,szab,kodama,hayward1,hayward2,adm_mass}. Notice that for the particular case of dust ($p=\PP=0$) we have $\MM =M$ and the QL mass--energy is conserved along the 4--velocity flow: $\dot \MM=0$. When $p,\,\PP$ are nonzero, then (\ref{QLM_mu}) generalizes the Friedman equation (\ref{fried_dust}). 

The function ``$\mu_*$'' that emerges from the integral relation between $\mu$ and $\MM$ in (\ref{QLM_mu}) motivates us to define the following map acting on any scalar function:\\

\noindent
{\underline{Definition. QL scalars.}} Let $X$ be the set of all smooth integrable scalar functions in a comoving regular domain $\DD=\mathbb{S}^2\times\xi\subset \T(t)$, where $\mathbb{S}^2$ is the unit 2--sphere parametrized by $(\theta,\phi)$ and $\xi =\{x\,|\, 0\leq x\leq r\}$, with $x=0$ marking a symmetry center. For any function $A\in X$, a ``dual'' QL scalar function follows from the map $\JJ_*: X\to X$ such that   
\begin{equation}A_* \equiv \JJ_*(A) =  \frac{\int_{x=0}^{x=r}{A\,R^2R' dx}}{\int_{x=0}^{x=r}{R^2R' dx}}.\label{QLfunc}\end{equation}
The QL scalar functions $A_*\,:\DD\to \mathbb{R}$ depend on the upper integration limit $r$ of $\xi$, and generalize to any scalar the QL mass--energy definition (\ref{QLM_mu}) constructed with $\mu$. See \cite{QLvars} for a more detailed discussion and a covariant definition of the integral (\ref{QLfunc}).\\

QL scalars $A_*$ defined by (\ref{QLfunc}) comply with the following useful properties:
\bse\label{props}\ba A - A_* &=& \frac{1}{R^3(r)}\int_{0}^{r}{A' R^3 dx},\label{prop1}\\
    A_*{}' &=& \frac{3R'}{R}\,[A-A_*],\label{prop2}\\
\dot A_* &=& (\dot A)_*+3(\HH A)_*-3\,\HH_*\, A_*,\label{prop3}\ea\ese 
where, in order to simplify notation, we have adopted the following conventions:
\ba \dot A_* &\equiv& \left[\JJ_*(A)\right]\,\dot{}\ne \JJ_*(\dot A)=(\dot  A)_*,\nonumber\\
 A_*{}' &\equiv& \left[\JJ_*(A)\right]\,' \ne \JJ_*(A') = (A')_*,\nonumber\\
 (AB)_* &\equiv& \JJ_*(AB),\nonumber\\
 \int_0^r{..\, \dd x} &\equiv& \int_{x=0}^{x=r}{\,..\, \dd x},\nonumber\\\label{notcon}\ea
It is straightforward to verify that these properties are perfectly self--consistent. 

From (\ref{theta1}), (\ref{RR1}) and (\ref{QLfunc}), the QL scalars dual to $\HH,\,\KK$ are
\bse\label{HKql}\ba \HH_* &=& \JJ_*(\HH)=\frac{\dot R}{R},\label{theta2}\\
 \KK_* &=& \JJ_*(\KK)=\frac{K}{R^2}\quad\Rightarrow\quad \frac{\dot\KK_*}{\KK_*}=-2\HH_*.\label{RR2}
 \ea\ese 
Hence, by applying (\ref{QLfunc}) to (\ref{mu1}) and (\ref{p1}) we get
\bse\ba \kappa\int_0^r{\mu R^2R'dx} &=& \frac{\kappa}{3}\mu_*R^3=R\left(\dot R^2+K\right),\\
\kappa\int_0^r{p R^2R'dx} &=& \frac{\kappa}{3}p_*R^3=-\frac{1}{3}\left[R\left(\dot R^2+K\right)+2\ddot R R\right],\nonumber\\\ea\ese
so that, with the help from (\ref{HKql}), these equations become identical to the Raychaudhuri equation of FLRW models, and its integral, the  Friedman equation:
\ba \dot\HH_* &=& -\HH_*^2-\frac{\kappa}{6}\,(\mu_*+3p_*),\label{Raych2}\\
\HH_*^2  &=& \frac{\kappa}{3}\,\mu_*-\KK_*,\label{cHam2}\ea
where $\HH_*^2 =[\JJ_*(\HH)]^2 \ne \JJ_*(\HH^2)$. These equations can be combined to yield the FLRW energy balance equation
\begin{equation}  \dot\mu_* = -3\,(\mu_*+p_*)\,\HH_*.\label{Econs2}\end{equation}
Applying (\ref{QLfunc}) to (\ref{PP1}), (\ref{Sigma1})  and (\ref{EE1}) yields the remaining scalars, $\Sigma$,\,$\PP$ and $\EE$, as deviations of $\HH,\,\mu$ and $p$ from their QL duals
\bse\label{PSE2}\ba \Sigma &=& -[\HH-\HH_*],\label{Sigma2}\\
\PP &=& \frac{1}{2}\,[p-p_*],\label{PP2}\\
\EE &=& -\frac{\kappa}{6}\,\left[\mu-\mu_* +\frac{3}{2}(p-p_*)\right],\label{EE2}\ea\ese
while the QL mass--energy function complies with
\begin{equation} 2\MM =\frac{\kappa}{3} \mu_* R^3,\qquad 2\dot\MM = -\kappa \,p_* R^2\dot R. \label{QLM3}\end{equation}

\section{Evolution equations for the QL scalars.}\label{eveqs}

The local and QL scalars can be related by means of ``relative deviations''
\begin{equation} \Da \equiv \frac{A-A_*}{A_*},\quad \Rightarrow\quad A = A_*\,\left[1+\Da\right].\label{Da_def}\end{equation}
which allows us to eliminate $\mu,\,p,\,\HH$ in terms of their duals $A_*$ and the corresponding $\Da$. Hence, bearing in mind (\ref{PSE2}), we have a complete scalar representation of LTB spacetimes given by
\begin{equation} \{\HH_*,\,\mu_*,\,p_*,\,\KK_*,\,\Dh,\,\Dm,\,\Dp,\,\Dk\}.\label{ql_scals}\end{equation}
which is alternative to the local representation (\ref{loc_scals}). We will denote (\ref{ql_scals}) the ``QL scalar representation''. The 1+3 system (\ref{ev_theta_13})--(\ref{cHam_13}) is the set of evolution and constraint equations for (\ref{loc_scals}), we derive now the evolution and constraint equations for the representation (\ref{ql_scals}).    

From (\ref{prop2}), the radial gradients of $\mu_*,\,p_*$ and $\HH_*$ can be given interms of the $\delta$ functions
\begin{equation} \frac{\HH_*{}'}{\HH_*} = \frac{3R'}{R}\,\Dh,\qquad \frac{\mu_*{}'}{\mu_*} = \frac{3R'}{R}\Dm,\qquad \frac{p_*{}'}{p_*} = \frac{3R'}{R}\Dp, \label{rad_grads}\end{equation}
while (\ref{Raych2}) and (\ref{Econs2}) are evolution equations for $\mu_*$ and $\HH_*$. Hence, the evolution equations for $\Dm$ and $\Dth$ follow from the consistency condition
\begin{equation} \left[A_*{}'\right]\,\dot{}=\left[\dot A_*\right]',\end{equation}
applied to (\ref{Raych2}), (\ref{Econs2}) and (\ref{rad_grads}) for $A=\HH_*,\,\mu_*$. The result is the following set of autonomous evolution equations for the QL representation (\ref{ql_scals}):
\bse\label{eveqs_ql}\ba 
\dot\mu_* &=& -3\,\left[\,1+w\,\right]\,\mu_*\,\HH_*,\label{evmu_ql}\\
\dot\HH_* &=& -\HH_*^2 -\frac{\kappa}{6}\,\left[\,1+3\,w\,\right]\,\mu_*,
\label{evHH_ql}\\
\dDm &=& 3\HH_*\,\left[\left(\Dm-\Dp\right)\,w-\left(1+w+\Dm\right)\Dh\right],
\nonumber\\
\label{evDmu_ql}\\
\dDh &=& -\HH_*\,\left(1+\Dh\right)\,\Dh\nonumber\\
 &{}&\,\,+ \frac{\kappa\mu_*}{6\,\HH_*}\left[\Dh-\Dm+3w\,\left(\Dh-\Dp\right)\right],\nonumber\\
\label{evDth_ql}
\ea\ese
where
\begin{equation}w\equiv \frac{p_*}{\mu_*}.\label{wdef}\end{equation}
The spacelike constraints associated with these evolution equations are simply the spatial  gradients (\ref{rad_grads}), while the Friedman equation (or Hamiltonian constraint) is (\ref{cHam2}). Equations (\ref{evmu_ql})--(\ref{evDth_ql}) become fully determined once an ``equation of state'' that fixes $p_*,\,\Dp$ as functions of $\mu_*,\,\Dm$ is selected (see section VII).  

With the help of (\ref{PSE2}) and (\ref{Da_def}), it is straightforwards to  prove that the evolution equations (\ref{eveqs_ql}) and the constraints (\ref{cHam2}) and (\ref{rad_grads}) are wholly equivalent to the 1+3 evolution equations (\ref{eveqs_13}) and their constraints (\ref{c_13}). Hence, given an ``equation of state'', they completely determine the dynamics of LTB spacetimes.

\section{A non--linear perturbation scheme}\label{perturb}

The definition (\ref{Da_def}) and the evolution equations (\ref{evmu_ql})--(\ref{evDth_ql}) suggest that the role of the functions $\Da$ can be rigorously characterized as  spherical perturbations on a formal FLRW ``background'' state given by the $A_*$. The perturbation formalisms developed by  Ellis, Bruni, Dunsbury and van Elst~\cite{ellisbruni89,BDE,LRS,1plus3} and Bardeen~\cite{bardeen} define a perturbation scheme in terms of a suitable map $\Phi$ between objects $\bar A$ in $\bar S$ (a FLRW model) and objects $A$ in $S$ (the ``perturbed'' lumpy model). The perturbation then compares $A$ with $\Phi(\bar A)$, which are objects in $S$ that define a formal FLRW ``background''. The application of this formalism to the case when the lumpy model is an  LTB spacetime simplifies considerably because (i) both classes of spacetimes (LTB and FLRW) can be completely described by covariant scalars, and (ii) both are given in the same normal comoving coordinate (or frame) representation. Hence, $\Phi$ can be a map between covariant scalars in $\bar S$ and covariant scalars $S$, and practically all problems due to gauge and coordinate freedom disappear. 

The map $\Phi$ can be defined as follows: let $\bar X$ and $X$ be, respectively, the sets of smooth integrable scalar functions in $\bar S$  and $S$, then for all covariant FLRW  scalars $\bar A\in \bar X$ (we denote FLRW objects with an over--bar), $\Phi$ is the map
\begin{equation} \Phi: \bar X\to X,\qquad \Phi(\bar A)=\JJ_*(A)=A_*\in X,\label{pert_map}\end{equation}
which characterizes QL scalars $\{\mu_*,\,p_*,\,\HH_*,\,\RR_*\}$ (LTB objects satisfying FLRW dynamics) as the ``background'' model in LTB spacetimes, while the perturbations $\Da$ provide their comparison with covariant scalars $A$
\begin{equation}  \Da = \frac{A-\Phi(\bar A)}{\Phi(\bar A)}.\label{pert_Da}\end{equation}

Following Dunsbury, Ellis and Bruni~\cite{ellisbruni89,BDE}, a perturbation scheme on FLRW cosmologies is covariant if $S$ is described by the 1+3 fluid flow variables, as in (\ref{eveqs_13}) \cite{ellisbruni89,BDE,LRS,1plus3}. Although our description of LTB spacetimes is not based on these scalars (the representation (\ref{loc_scals})), it is still covariant because $\{\mu_*,\,p_*$ and $\Theta_*\}$, are themselves covariant scalars by virtue of their connection with the invariants $\MM,\,R$ and their derivatives in (\ref{HKql}) and (\ref{QLM3}) (see \cite{kodama}). Hence, the formalism associated with (\ref{pert_map}) and (\ref{pert_Da}) is covariant. 

Also, by virtue of the Stewart--Walker gauge invariance lemma \cite{ellisbruni89}, all covariant objects in $S$ that would vanish in the background $\bar S$ (a FLRW cosmology in this case) are gauge invariant (GI), to all orders, and also in the usual sense (as in \cite{bardeen}). The background variables $\mu_*,\,p_*,\,\Theta_*$ do not vanish for $\bar S$, hence they are ``zero order'' GI variables to all orders. The  quantities in LTB spacetimes that vanish for a FLRW cosmology are the tensors $\Pi^{ab},\,\sigma^{ab}$ and $E^{ab}$, given by (\ref{PSEsc}) in terms of the scalar functions $\PP,\,\Sigma$ and $\EE$ in (\ref{PP1})--(\ref{ESsc}). But from (\ref{PSE2}), these functions are basically the fluctuations $\mu-\mu_*$,\, $p-p_*$ and $\Theta-\Theta_*$.  Hence, from (\ref{Da_def}), (\ref{rad_grads}) and (\ref{pert_Da}), the perturbation variables $\Dm,\,\Dp$ and $\Dth$, as well as the gradients $\mu_*',\,p_*'$ and $\Theta_*'$,  are all ``first order'' quantities that are GI to all orders. Therefore, LTB spacetimes in the QL scalar representation $\{A_*,\,\Da\}$ can be rigorously characterized as spherical, non--linear GIC perturbations on a FLRW background. In the linear limit $|\Da|\ll 1$ these perturbations reduce to spherical perturbations in the long wavelength approximation and in the synchronous gauge~\cite{QLvars}.

\section{Equations of state.}\label{EOS}

The system (\ref{eveqs_ql}) is still undetermined, as there are no evolution equations for $p_*$ and $\Dp$. As in any perturbative approach, we need to impose an ``equation of state'' (EOS) between the background variables, which are in this case $p_*$ and $\mu_*$. Such an EOS determines the ``background subsystem'' (\ref{evmu_ql})--(\ref{evHH_ql}), as well as  the relation between the perturbations $\Dm$ and $\Dp$.

Consider a commonly used barotropic equation of the form
\begin{equation} p_* = p_*(\mu_*).\label{BEOS}\end{equation} 
From (\ref{rad_grads}), we obtain the corresponding relation between fluctuations and perturbations of $\mu$ and $p$
\bse\label{BEOS_perflu}\ba p-p_* &=& \frac{d p_*}{d \mu_*}\,[\,\mu-\mu_*\,],\label{BEOSfluc}\\
\Dp &=& \frac{d\ln p_*}{d\ln\mu_*}\,\Dm.\label{BEOSpert}\ea\ese
Notice that, in general, the EOS (\ref{BEOS}) will not be satisfied by $p$ and $\mu$. But this is expected, since in a perturbation scheme fluctuations or perturbations do not satisfy, in general, the EOS of the background (not even in linear perturbations). The exception is the simple particular case of (\ref{BEOS}) 
\begin{equation} p_* = w_0\,\mu_*\,,\label{BEOSlin}\end{equation} 
where $w_0=w_0(t)$, and in particular $w_0$ can be a constant (so that $w_0=0$ is dust). The EOS (\ref{BEOSlin}) leads to
\begin{equation}p-p_* = w_0\,[\,\mu-\mu_*\,],\qquad
\Dp = \Dm.\label{BEOSlinFP}\end{equation}
and so $p=w_0\,\mu$, even if $w_0=w_0(t)$ because the map $\JJ_*$ in (\ref{QLfunc}) involves integrals of scalars along hypersurfaces of constant $t$, hence $\JJ_*(w_0(t),\,\mu)=w_0(t)\,\JJ_*(\mu)$ for all $t$.  

It could be argued that, because the perturbations are ``exact'' and non--linear ({\it i.e.} $\Da$ can be ``large''), even if a given EOS (\ref{BEOS}) is physically reasonable we could have a wholly unphysical relation between $p$ and $\mu$ through (\ref{BEOSfluc}). While this possibility cannot be ruled out, it is something that must be tested by looking at the solutions of (\ref{eveqs_ql}) for different EOS. Evidently, the plausibility of the formalism must be verified and judged for different EOS according to the predictions of the solutions of (\ref{eveqs_ql}). 

However, it would be wholly incorrect to disregard the formalism from the outset on the grounds that local variables $\mu$ and $p$ do not satisfy a given ``familiar'' EOS (say, the Chaplygin gas) used in a FLRW framework. Not only we are treating $\mu$ and $p$ as non--linear perturbations (not expected to satisfy the EOS of the background), but also relations that could be valid in a FLRW context (where local and QL variables are identical) need not hold automatically in an inhomogenous non--linear context. Demanding that the local EOS must be considered as the only physically meaningful EOS in an inhomogeneous context is  only correct for hydrodynamical sources characterized by short range interactions or for self--gravitating systems in a kinetic regime \cite{Padma_GTD}. There is no reason to impose such condition in self--gravitating systems dominated by long range interactions.  

As shown in Newtonian self--gravitating systems in specific scales~\cite{saslaw1,saslaw2}, the correlation effects of the long range nature of gravity appear as a first order virial correction to the classical ideal gas EOS linking $\mu$ and $p$. By looking at (\ref{BEOSfluc}), it is evident that an EOS like (\ref{BEOS}) introduces similar correction terms to the relation between local variables $\mu$ and $p$ (effects of this type in the classical ideal gas are discussed in \cite{QLvars}). These corrections could convey important non--local information in the study of self--gravitating systems, as for example the so--called ``back--reaction'', which can be related to dynamical and observational effects of QL energy~\cite{theo1,theo2,wiltshire}. Non--local correlation terms could also be important in looking at self--gravitating systems with relativistic dark energy sources, as there is no reason to assume that these sources (whose fundamental nature is still unknown) are of a hydrodynamical nature even under inhomogeneous and non--linear conditions.

\section{Mixtures of dark matter and dark energy.}\label{mixtures} 

\subsection{General mixtures} 

The formalism that we have presented can be extended to describe sources like fluid mixtures, for example, a mixture of dark matter (DM) and dark energy (DE). The resulting models provide an inhomogeneous generalization to known DM-DE mixtures in a  FLRW
context~\cite{intmix1,intmix2,intmix_new,EOS1,EOS2}. To examine a DM-DE mixture we assume a decomposition of the ``total'' energy--momentum tensor as a sum $T^{ab}=T_{_{\rm{DM}}}^{ab}+T_{_{\rm{DE}}}^{ab}$,
which implies that energy density and pressures split as 
\bse\label{Tabdec}\ba \mu &=& \mu_{_{\rm{DM}}} + \mu_{_{\rm{DE}}},\\ p &=& p_{_{\rm{DM}}} +
p_{_{\rm{DE}}},\\
\Pi^{ab}&=&\PP\Xi^{ab} = [\PP_{_{\rm{DM}}}+\PP_{_{\rm{DE}}}]\,\Xi^{ab},\ea\ese
where $\Xi^{ab}$ was defined in (\ref{PSEsc}). In order to simplify notation  we introduce:
\ba\rho &=& \mu_{_{\rm{DM}}},\qquad \varphi= p_{_{\rm{DM}}},\nonumber\\
    q &=& \mu_{_{\rm{DE}}},\qquad \pi=p_{_{\rm{DE}}},\nonumber\\
        \mu_* &=& \rho_*+q_*,\qquad p_* = \varphi_*+\pi_*,\label{newnames}\ea
while, from (\ref{PP2}), $\PP$ splits as in (\ref{Tabdec})
\begin{equation}\PP_{_{\rm{DM}}} =\frac{\varphi-\varphi_*}{2},\qquad \PP_{_{\rm{DE}}}   =\frac{\pi-\pi_*}{2}.\label{PPsplit}\end{equation}
The total conservation law $\nabla_bT^{ab}=0$ can be split in conservation laws for  the
individual tensors 
\begin{equation}\nabla_b T_{_{\rm{DM}}}^{ab}= j^a=-\nabla_b T_{_{\rm{DE}}}^{ab},
\end{equation}\
where $j^a$ is the interaction current (or coupling) that characterizes an interactive mixture,  so that if $j^a=0$ the mixture is non--interactive (decoupled).  We take this current as a vector parallel to the
4--velocity, so that $j_a=Ju_a$ and $h_{ca}j^a=0$ hold. The spatially projected conservation
equation $h_{ac}\nabla_bT^{ab}=0$ remains as it is, but the projection along $u^a$ becomes
\begin{equation}u_a \nabla_b T_{_{\rm{DM}}}^{ab}= J =- u_a \nabla_b T_{_{\rm{DE}}}^{ab}.
\label{uTab_cons_J}\end{equation}
which, from (\ref{ev_mu_13}), (\ref{Tabdec}) and (\ref{newnames}), takes the form
\bse\label{local_cons}\ba \dot\rho+3(\rho+\varphi)\HH-3(\HH-\HH_*)(\varphi-\varphi_*)=J,\\
\dot q+3(q+\pi)\HH-3(\HH-\HH_*)(\pi-\pi_*)=-J,\ea\ese
where we used (\ref{PSE2}) and (\ref{PPsplit}) to eliminate $\PP$ and $\Sigma$  in terms of the
fluctuations of $\varphi,\,\pi$ and $\HH$. With the help of (\ref{prop3}) and the relation between local and quasi--local variables (\ref{Da_def}), and considering that 
\begin{equation} J=J_*\,[1+\delta^{(J)}],\label{deltas_mixt}\end{equation}
we can transform (\ref{eveqs_ql}) into the following set of evolution equations for an
interactive mixture in the QL scalar representation:
\begin{widetext}\bse\label{eveqs_mixt}\ba \dot\HH_*&=& -\HH_*^2
-\frac{\kappa}{6}\left[\rho_*(1+3w_\rho)+q_*(1+3w_q)\right],\label{eveq_mixt_HH}\\
 \dot\rho_* &=& -3\,\rho_*\,[1+w_\rho]\,\HH_*+J_*,\label{eveq_mixt_rho}\\
  \dot q_* &=& -3\,q_*\,[1+w_q]\,\HH_*-J_*,\label{eveq_mixt_q}\\
   \dDrho &=& 3\HH_*\,[w_\rho\,(\Drho-\delta^{(\varphi)})
-(1+w_\rho+\Drho)\Dh]-\frac{J_*}{\rho_*}(\Drho-\delta^{(J)}),\\
 \dDq &=& 3\HH_*\,[w_q\,(\Dq-\delta^{(\pi)})-(1+w_q+\Dq)\Dh] -\frac{J_*}{q_*}(\Dq-\delta^{(J)}),\\
 \dDh &=& -\HH_*\Dh(1+\Dh)+\frac{\kappa}{6\HH_*}
\left[\rho_*\,(\Dh-\Drho)+q_*\,(\Dh-\Dq)
+3\varphi_*\,(\Dh-\delta^{(\varphi)})+3\pi_*\,(\Dh-\delta^{(\pi)})\right],\nonumber\\
\ea\ese\end{widetext}
where
\begin{equation}w_\rho\equiv \frac{\varphi}{\rho},\qquad w_q\equiv 
\frac{\pi}{q}.\end{equation}
The evolution equations for a non--interactive mixture follow by  setting
$J_*=\delta^{(J)}=0$. An important particular case follows by considering DM as cold dark matter (CDM),
hence  $T_{_{\rm{DM}}}^{ab}=T_{_{\rm{CDM}}}^{ab}$ takes the form of dust
($\varphi=\varphi_*=0$). We examine numerically an interactive mixture in section \ref{Imixture}. 

\subsection{Scalar fields.}

The scalar field (``quintessense'') is a very popular model for dark energy sources. Unfortunately, the energy--momentum tensor of an inhomogeneous scalar field is incompatible with the LTB metric. Hence, the only way in which we can model a scalar field with LTB spacetimes is by considering it as homogeneous DE, while DM  is then an inhomogeneous fluid that can be (if desired) CDM dust, but this is not a necessary condition, though if we take DM as CDM, then $T^{ab}$ takes the form of a perfect fluid ($\PP=0$). 

The particular case of a mixture with homogeneous DE follows from (\ref{eveqs_mixt}) with $\Dq = \delta^{(\pi)}=0$, so that $q=q_*$ and $\pi=\pi_*$, and it implies the following constraint on the interaction term $J_*$ 
\begin{equation}J_*'=3q_*(1+w_q)\HH_*'\qquad\Rightarrow\quad J_*\delta^{(J)}=3q(1+w_q)\HH_*\Dh,\end{equation}
so that uncoupled mixtures are not possible ($J_*=0$ implies a FLRW spacetime $\Dh=0$). In particular, we can always assume a scalar field form with arbitrary potential for the DE homogeneous fluid: 
\begin{equation} q = \frac{\dot\phi^2}{2}+V(\phi),\qquad \pi = \frac{\dot\phi^2}{2}-V(\phi),\end{equation}
with $\phi=\phi(t)$, which transforms (\ref{eveq_mixt_q}) into the Klein--Gordon equation
\begin{equation} \ddot\phi+3\HH_*\dot\phi\left(1+\frac{\Dh}{\delta^{(J)}}\right)+\frac{dV}{d\phi}=0,\end{equation}
The mixture examined in \cite{intmix3} is the particular spatially flat case of such a mixture, in which the interaction term $J_*$ was obtained from the Lagrangian of a scalar--tensor theory.

\section{A FLRW--like LTB metric.}\label{FLRWlike}

It is useful for the numerical solution of the evolution equations to re--parametrize the LTB metric (\ref{LTB}) so that it looks as close as possible to a FLRW line element. This metric is invariant under arbitrarily re--scaling of the  radial coordinate $r=r(\tilde r)$, hence we will use this gauge freedom to chose this coordinate so that
\begin{equation} R_i \equiv R(ct_i,r)= h_s^{-1}\,r,\label{Yi1}\end{equation}
where $h_s^{-1}$ is an arbitrary characteristic length scale. Considering (\ref{Yi1}), we introduce the following dimension--less form for the metric functions $R$ and $R'$:
\bse\label{LG}\ba \ell &\equiv& \frac{R}{R_i}=\frac{R}{ h_s^{-1} r}\qquad 
\Rightarrow\qquad \ell_i=1,\label{Ldef}\\
    \Gamma &\equiv& \frac{R'/R}{R_i'/R_i}=1+\frac{r\ell'}{\ell},\quad    
\Rightarrow\quad \Gamma_i=1,\label{Gdef}\ea\ese
where the subindex ${}_i$ will henceforth denote evaluation at a fiducial  hypersurface $\T_i$ marked by $t=t_i$, which can be considered as an initial surface for treating the integration of (\ref{eveqs_ql}) and (\ref{eveqs_mixt}) as a well posed
initial value problem. Notice that $[A_*]_i=[A_i]_*$ holds in general. We can now write (\ref{rad_grads}) for any $A_*$ as
\begin{equation}\frac{A_*'}{A_*}=\frac{3\Gamma}{r}\,\Da.\label{rad_grads2}\end{equation}

The parametrization given by (\ref{Yi1}) and (\ref{LG}) transforms the LTB metric  (\ref{LTB}) into the following form that resembles a FLRW line element
\begin{equation}ds^2=-c^2dt^2+L^2\, \left[\frac{\Gamma^2\,dr^2}{1-[k_*]_i r^2}
+r^2\,(d\theta^2+\sin^2\phi)\right],\label{LTB2}\end{equation}
where  
\begin{equation}L \equiv h_s^{-1}\ell, \qquad [k_*]_i\equiv \frac{[\RR_*]_i}{6\,h_s^2}
=\frac{[\KK_*]_i}{h_s^2}.\label{Lki_def}\end{equation}
In terms of $\ell$,  the QL scalars in (\ref{HKdefs}) and (\ref{HKql}) take the  following FLRW--like forms
\ba \HH_* &=& \frac{\Theta_*}{3}=\frac{\dot \ell}{\ell}=\frac{\dot L}{L},
\label{ave_L}\\        \KK_* &=&
\frac{\RR_*}{6}=\frac{[\KK_*]_i}{\ell^2} = \frac{[k_*]_i}{L^2},\label{ave_K}\ea
which highlights the role of $\ell$ (or $L$) as a position dependent scale factor,  as opposed to $R$, whose geometric meaning is the ``area distance'' or curvature radius of the orbits of
SO(3).

The radial coordinate choice associated with $R_i$ in (\ref{Yi1}) is appropriate when we have a symmetry center marked by $r=0$, but then it is perfectly valid to set the radial coordinate so that
\begin{equation}R_i = h_s^{-1} f(r),\label{Yi_gen}\end{equation}
where $f(r)$ is any smooth positive and monotonously increasing function so that $f(0)=0$. Such choice was made in the mixture example of section \ref{Imixture} (see Appendix D). Other choices of $f(r)$ depend on the topological class of the hypersurfaces $\T$, for example, if they have $\mathbb{S}^3$ topology (like a ``closed'' FLTW cosmology), then a good choice for $f(r)$ would be a sine--like function with a second zero at the second symmetry center. For wormhole topologies, $\mathbb{S}^2\times \mathbb{R}$ or $\mathbb{S}^2\times \mathbb{S}^1$, the function $f$ should not have zeroes but must have at least a ``turning value'' where $f'=0$. See \cite{suss08,ltbstuff,suss02}.

\section{The ``homogeneous subsystem'' and regularity conditions. }\label{FLRWsubs}

The evolution equations (\ref{evmu_ql}) and (\ref{evHH_ql}) for the background QL variables $\mu_*,\,\HH_*$ are formally identical to FLRW evolution equations (like the evolution equations for  $\rho_*,\,q_*,\,\HH_*$ in (\ref{eveqs_mixt})). This suggests exploring the possibility of integrating separately these ``homogeneous subsystems'', or even using known FLRW solutions, to determine analytically the background variables in (\ref{eveqs_ql}) and (\ref{eveqs_mixt}). 

The metric form (\ref{LTB2}), with (\ref{ave_L}) and (\ref{ave_K}), allows us to re--write the evolution equations (\ref{evmu_ql})--(\ref{evHH_ql}) and the Friedman equation (\ref{cHam2}) in the following familiar FLRW form: 
\bse\label{fried1}\ba
\frac{\ddot \ell}{\ell} &=& -\frac{\kappa}{6}\,\left[\mu_*+3p_*\right],\label{fried10}\\ 
\frac{3\dot \ell}{\ell} &=& -\frac{\dot\mu_*}{\mu_*+p_*},
\label{fried11}\\
\dot \ell^2 &=& \frac{\kappa}{3}\mu_*\,\ell^2-[\KK_*]_i.\label{fried12}\ea\ese
with $\ell(t,r)$ playing the role of a position--dependent dimension-less FLRW scale factor $a(t)/a(t_0)$ and $[\KK_*]_i$ is a radially dependent ``curvature index''.  Once an EOS like (\ref{BEOS}) is given, equation (\ref{fried11}) can be formally integrated leading to
\begin{equation}\mu_* = \mu_*(\ell,[\mu_*]_i),\label{muav_func}\end{equation}   
where $[\mu_*]_i=\mu_*(t_i,r)$ is the initial QL energy density (the ``arbitrary constant''). Once (\ref{muav_func}) is known, the Friedman equation (\ref{fried12}) can be, in principle (and for simple enough EOS), integrated or reduced to quadratures~\cite{hydro}. 

While considering the integration of the FLRW subsystem (\ref{fried1}) (or using known solutions of it) is useful in itself, given the vast amount of known FLRW solutions \cite{review,chaplygin1,chaplygin2,intmix1,intmix2,intmix_new}, the resulting integrals are far from a solution to (\ref{eveqs_ql}) and (\ref{eveqs_mixt}).  As opposed to the FLRW case, (\ref{fried1})  are PDE's (not ODE's) and so initial values are functions of $r$ (not constants), hence there is a non--trivial radial dependence of all variables that must be still determined by solving the evolution equations for the $\Da$, which relate to radial gradients of the $A_*$ through (\ref{rad_grads}) and convey the departure from homogeneity. Still, the solutions to this subsystem are useful to discuss and understand a basic regularity condition: absence of shell crossings, which is also important in LTB dust solutions~\cite{kras,suss08,ltbstuff,suss02}. 

Since $\mu_*$ in (\ref{muav_func}) is, in general, independent of $\Gamma$, we can use use (\ref{Gdef}) and to eliminate $\ell'$ in terms of $\Gamma$ and applying (\ref{rad_grads2}) to this form of $\mu_*(\ell,[\mu_*]_i)$ we obtain the general relation
\begin{equation} \Dm =
\frac{1}{3\Gamma}\,\left[(\Gamma-1)\frac{\partial\ln\mu_*}{\partial\ln\ell}+3\Dim\,\frac{\partial\ln\mu_*}{\partial\ln[\mu_*]_i}\right],\label{Dm_ana}\end{equation}
It is evident from (\ref{Dm_ana}) that $\Dm$ and $\mu$ diverge for finite
$\mu_*$ if $\Gamma\to 0$ for $\ell>0$. Therefore, a necessary and sufficient condition to prevent this unphysical feature is to demand that initial conditions are selected, so that the following constraint:
\begin{equation} \Gamma(t,r) > 0,\label{no_xcr}\end{equation}
holds for all $(t,r)$ for which $\ell(t,r)>0$ (the range of physical evolution). This is equivalent to demanding absence of shell--crossings associated with $\Gamma=0$ occurring in the physical evolution range~\cite{suss08,ltbstuff,suss02}. This regularity condition applies to all scalars, such as $\HH_*$ and $\Dh$, and in general prevents an unphysical evolution in which perturbations $\Da$ diverge, hence making local scalars $A$ diverge, all of it while their associated QL variables $A_*$ remain finite. 

Initial conditions that fulfill (\ref{no_xcr}) can be given analytically for dust sources~\cite{ltbstuff,suss02,hela}. In general, this regularity condition must be verified numerically for any choice of initial conditions.

\section{Smooth and relativistic ``top hat'' models. }\label{tophats}

When assuming the existence of a symmetry center, a type of model that is often considered is that of a single spherical clump or void region that could be, somehow, embedded in an asymptotic homogeneous FLRW background. The non--linear perturbation scheme that we have presented is ideally suited for this type of scenarios, and it does not restrict the different ways in which boundary conditions can be set for this asymptotic blending into a cosmic background. 

One possibility for setting boundary conditions is simply to perform a smooth match, at a fixed comoving radius $r=r_{\rm b}$, between a region of an LTB spacetime (containing a center) and a suitable or ``equivalent'' FLRW spacetime (the precise meaning of ``equivalent'' will become clear further ahead). The resulting compound spacetime is formed by a comoving LTB region (a clump or void) in the range $0\leq r\leq r_b$, smoothy matched to a homogeneus FLRW section that extends for $r>r_{\rm b}$. This construction generalizes to smooth and relativistic conditions the so called Newtonian ``top hat'' or ``spherical collapse models'' that are widely used in astrophysical literature~\cite{tophats}, and that tend to overlook the details of the embedding of the overdensity or void into the FLRW background.

The section of a FLRW spacetime to be matched with a LTB comoving region is described by the metric
\begin{equation}ds^2=-c^2dt^2+a^2(t)\, \left[\frac{dr^2}{1-k_0 r^2}+r^2\,
(d\theta^2+\sin^2\phi)\right],\label{FRLWmetric}\end{equation}
where $k_0=0,\pm 1$, and $r> r_{\rm b}$. This FLRW section is characterized by the covariant scalars  $\{\bar\mu,\, \bar p,\,\bar\HH,\,\bar\KK\}$, with $\bar \HH=\dot a/a$ and $\bar\KK=k_0/a^2$ (FLRW scalars are denoted by an overbar). The metric of the LTB region is just (\ref{LTB2}) with $0\leq r\leq r_{\rm b}$. Since both metrics are already given in the same normal comoving coordinates, the conditions for a smooth matching between them simplify considerably.  

The standard matching conditions between two spacetimes are those of Darmois~\cite{ltbstuff,darmois}. For the LTB--FLRW case,  with the metrics (\ref{FRLWmetric}) and (\ref{LTB2}), these conditions imply~\cite{ltbstuff}  continuity at $r_{\rm b}$ of their metric function $g_{\theta\theta}$ and its proper time derivative (which is coordinate time). From (\ref{LG}), (\ref{LTB2}), (\ref{ave_L}), (\ref{ave_K})  and (\ref{FRLWmetric}), these requirements lead to 
\bse\label{Darmois1}\ba L_{\rm b} &=& a(t),\label{Darmois11}\\
 \left[\HH_*\right]_{\rm b} &=& \bar\HH(t), \label{Darmois12}\\
  \left[\mu_*\right]_{\rm b} &=& \bar\mu(t),\label{Darmois13}\\
    \left[k_*\right]_i(r_{\rm b}) &=& k_0,\label{Darmois14}\ea\ese 
where the subindex ${}_{\rm b}$ indicates evaluation at $r=r_{\rm b}$.   

For a tensor $T^{ab}$ with  anisotropic stresses like (\ref{Tab}), Darmois conditions also imply \cite{darmois} the continuity of the total pressure in the radial direction, $T^r\,_r=p-2\PP$, hence besides (\ref{Darmois1}), we have from
(\ref{PP2})
\begin{equation}[p-2\PP]_{\rm b} = [p_*]_{\rm b} =\bar p(t).\label{Darmois2}\end{equation}
 Since in the FLRW region QL and local scalars are identical: \, $\bar p=\bar p_*$, in the LTB  region we must have $[p-p_*]_{\rm b}\to 0$ as $r\to r_{\rm b}$ for all $t$. Hence, (\ref{PP2}), (\ref{rad_grads}) and (\ref{Darmois2}) clearly imply
\begin{equation}\PP_{\rm b}=0,\qquad p_*'(t,r_{\rm b})=0,\qquad [p-p_*]_{\rm b}=[\Dp]_{\rm b}=0,
\label{Pbzero}\end{equation}
so that (\ref{Tab}) takes the perfect fluid form at $r=r_{\rm b}$. 

Since Darmois matching conditions require continuity of the QL scalars $\{\mu_*,\,p_*,\,\KK_*\}$ and the FLRW scalars $\{\tilde\mu,\,\tilde p,\,\tilde\KK\}$,  the ``equivalent'' FLRW spacetime to be smoothly matched to a given LTB region can be precisely defined as that in which $\bar p$ and $\bar \mu$ satisfy the same EOS that $p_*$ and $\mu_*$ satisfy in the LTB region. Also, from (\ref{ave_K}), (\ref{Darmois11}) and (\ref{Darmois14}), QL spatial curvature must be continuous: $\left[\KK_*\right]_{\rm b}=k_0/a^2(t)$. Hence, the spatial curvature of the whole matched FLRW region is determined by the sign of $\left[\KK_*\right]_{\rm b}$. Notice that  $\KK_*$ changing sign for values $r<r_{\rm b}$ in the LTB region has no effect on the matching.    

While Darmois matching conditions do not require continuity at $r=r_b$ of the  local energy density $\mu$, nor of the metric component $g_{rr}$, it is highly desirable to avoid discontinuity of these quantities. Hence, we supplement these conditions by further demanding that
\bse\label{extraMC}\ba \mu_{\rm b} &=& \left[\mu_*\right]_{\rm b}=\bar\mu(t)\quad\Rightarrow\quad 
\mu_*'(t,r_{\rm b})=0,\label{extraMC1}\\
         k_i'(r_{\rm b}) &=& 0,\label{extraMC2}\\
                  \Gamma_{\rm b} &=& 1.\label{extraMC3}\ea\ese 
 
The complete set of matching and regularity conditions (\ref{Darmois1})--(\ref{extraMC}) lead to a well defined and self--consistent model of a spherical comoving LTB region that is smoothly matched to a appropriate FLRW background. These conditions imply:
\begin{equation} \left[\Dm\right]_{\rm b}=\left[\Dp\right]_{\rm b}=\left[\Dh\right]_{\rm b}=0,\end{equation}
and consequently their time derivatives vanish as well. The system (\ref{eveqs_ql}) reduces at $r=r_{\rm b}$ (and for $r> r_{\rm b}$) to the FLRW equations (\ref{evmu_ql})--(\ref{evHH_ql}), while the mixture equations (\ref{eveqs_mixt}) reduce in the same ranges to the FLRW equations (\ref{eveq_mixt_HH})--(\ref{eveq_mixt_q}).  

\section{Applications}\label{applications} 

We examine in this section two numerical applications of the perturbation formalism: black hole formation in a Chaplygin gas universe and evolution of voids in a mixture of cold dark matter (CDM) and dark energy (DE). These examples are idealized because their purpose is to illustrate how the formalism can work in practice. More ``realistic'' and comprehensive studies, and/or models with a careful fitting to obervations, would require separate articles in their own merit. 

We still need for the numerical solution of the evolution equations (\ref{eveqs_ql}) and (\ref{eveqs_mixt}) to re--cast the QL variables in dimension--less form (the perturbations $\Dm,\,\Dp,\,\Dh$ are already dimensionless).  Dimensionless quantities can be constructed by means of the freedom afforded by the arbitrary length scale $h_s^{-1}$ defined in (\ref{Yi1}):
\begin{equation}\mu_* \rightarrow \frac{\kappa\,\mu_*}{3 h_s^2},\qquad p_* \rightarrow
\frac{\kappa\,p_*}{3 h_s^2},\qquad \HH_* \rightarrow
\frac{\HH_*}{h_s},\label{dimensionless1}\end{equation}
which can also be done with the QL variables in (\ref{eveqs_mixt}). The evolution time parameter, $ct$, must then be replaced by the dimensionless time
\begin{equation}\tau \equiv h_s c(t-t_i).\label{dimensionless2}\end{equation}
so that $\tau=0$ corresponds to $\T_i$. The initial conditions and other technical issues connected to the numeric integration of evolution equations in this section are presented and discussed in the Appendices.

\subsection{Gravitational collapse in a Chaplygin gas universe}\label{chaplygin}

The Chaplygin gas can be studied as an LTB spacetime in a fully relativistic inhomogeneous context. So far, this source has been examined under inhomogeneous conditions only in a Newtonian context~\cite{chapinhom}, as non--linear but relativistic approximations~\cite{chapinhom2}, or as a static object~\cite{chapstar}.  

The LTB spacetime associated with the ``standard'' Chaplygin gas EOS~\cite{chaplygin1,chaplygin2} is that characterized by the following particular case of (\ref{BEOS})
\begin{equation}p_* = -\frac{\alpha}{\mu_*},\qquad w = -\frac{\alpha}{\mu_*^2},
\label{Chap_eqst}\end{equation}
where $\alpha$ is a constant. Following (\ref{BEOS_perflu}), the relation between pressure and density  fluctuations and perturbations is then
\bse\label{fluc_Chap}\ba p-p_* &=& 
\frac{\alpha}{\mu_*^2}\,(\mu-\mu_*),\\
     \Dp &=& -\Dm.\label{dpchap}\ea\ese
The evolution equations are simply (\ref{eveqs_ql}) specialized for the EOS  (\ref{Chap_eqst}) and
with $\Dp$ given by (\ref{dpchap}). Since $p=p_*[1+\Dp]$ and $\mu=\mu_*[1+\Dm]$, equations (\ref{fluc_Chap})  imply the relation
\begin{equation}p = -\frac{\alpha}{\mu}\,[1-(\Dm)^2],\label{Chap_local}
\end{equation}
which can also be interpreted as $p$ being expressible in terms of the familiar (local) EOS, with a first order virial correction due to squared fluctuations that convey the effect of long range interactions. This type of relation is qualitatively analogous to that between the state variables of the ideal gas under a self--gravitating regime in Newtonian systems~\cite{saslaw1,saslaw2} (see also \cite{QLvars}). As long as we ignore the fundamental physics of the Chaplygin gas, using the EOS (\ref{Chap_eqst}) and having the local variables $p,\,\mu$ given by (\ref{Chap_local}) in an inhomogeneous self--gravitating context, is a reasonable working hypothesis.     

The solutions to the homogeneous subsystem (\ref{fried1}) leads to the following forms of $\mu_*$  and $p_*$ as functions of $\ell$ and $[\mu_*]_i$
\bse\label{Chap_EPav}\ba \mu_* &=& \frac{\left\{\alpha\,(\ell^6-1)+
[\mu_*]_i^2\right\}^{1/2}}{\ell^3},\label{Chap_Eav}\\
p_* &=& -\frac{\alpha\,\ell^3}{\left\{\alpha\,(\ell^6-1)+[\mu_*]_i^2\right\}^{1/2}},
\label{Chap_Pav}\ea\ese
These equations clearly illustrate the well known behavior of the Chaplygin gas as dust--like  ($p_*\approx 0$) for
earlier times and as cosmological constant for latter times: 
\ba\mu_* &\approx& \frac{ \left\{ [\mu_*]_*^2-\alpha \right\}^{1/2} }{\ell^3},\quad p_*  \approx
\frac{-\alpha\,\ell^3}{ \left\{ [\mu_*]_*^2-\alpha \right\}^{1/2} },\quad\ell\ll 1\nonumber\\\label{Chap_asynt1}\\\mu_* &\approx& \sqrt{\alpha},\quad p_* \approx -\sqrt{\alpha},\qquad \ell\ll 1.
\label{Chap_asynt2}\ea
These asymptotic limits suggest considering the numerator of $\mu_*$ in (\ref{Chap_asynt1}) as a rest--mass density, while (\ref{Chap_asynt2}) suggest considering $\sqrt{\alpha}$ as a cosmological constant. Using the length scale $h_s^{-1}$ introduced in (\ref{Yi1}), we define the following dimensionless parameters 
\begin{equation} \lambda \equiv \frac{\kappa\sqrt{\alpha}}{3h_s^2},\qquad 
[m_*]_i \equiv \frac{\kappa}{3h_s^2} \left\{ [\mu_*]_i^2-\alpha \right\}^{1/2},\label{Chap_pars}\end{equation}
which lead to the dimension--less form for the initial value functions (see equations
(\ref{initc3})--(\ref{initc4}) in Appendix C). 

Applying the Chaplygin gas EOS to (\ref{Dm_ana}) and using (\ref{Chap_Eav}), we get the
following expression for $\Dm$
\begin{equation}\Dm =
\frac{(\ell^6\,\Gamma-1)\,\alpha+(1+\Dim)\,[\mu_*]_i^2}{\left\{(\ell^6-1)\,\alpha+[\mu_*]_i^2\right\}\,\Gamma}-1,\label{Dm_ana_ch}\end{equation}
which shows that $\Gamma\to 0$ for $\ell>0$ implies that $\mu$ and $p$ diverge for finite $\mu_*$ and $p_*$ (which do not depend on $\Gamma$, as can be seen from (\ref{Chap_EPav})). Hence, the condition (\ref{no_xcr}) to avoid shell crossings  must be imposed. In principle, it could be possible to use (\ref{Chap_Eav}) for solving the Friedman equation (\ref{fried12}), and use the solution to find $\Gamma$ (from \ref{Gdef})), and then $\Dm$ from (\ref{Dm_ana_ch}), however it is far easier (and more intuitive) to solve the numeric system (\ref{eveqs_ql}) for the Chaplygin gas EOS. 
 
We will consider a ``top hat'' model, as discussed in section \ref{tophats}, formed by an inhomogeneous Chaplygin gas comoving region, matched at a comoving boundary $r=r_{\rm b}$ with a Chaplygin gas FLRW background (its equivalent FLRW model). Assuming the initial inhomogeneity as that of an early universe small over--density with positive but small scalar curvature, expecting that comoving layers in the overdensity will expand and then re--collapse while the FLRW background keeps expanding. 

Since initial conditions depend on the parameters $[m_*]_i,\,\lambda$ and $[k_*]_i$, and we are assuming that $t=t_i$ ($\tau=0$) corresponds to nearly homogeneous early universe initial conditions, then we can take  $h_s = \tilde H_i$, where $ \tilde H_i$ is a FLRW Hubble length at $t=t_i$. Hence, by identifying $\Lambda=\sqrt{\ACal}$ and bearing in mind that in the present era $t=t_0$ we have
\begin{equation}\Omega_0^{\Lambda}=\frac{\kappa\,\Lambda}{3\tilde H_0^2}=\lambda\,
\left(\frac{\tilde H_i}{\tilde H_0}\right)^2\approx 0.7,\end{equation} 
thus, $\lambda\sim 10^{-3}\ll 1$ since $\tilde H_i\gg \tilde H_0$. On the other hand, nearly homogeneous and spatialy flat initial conditions imply $[k_*]_i\approx 0$ and $[m_*]_i \approx  1$.  The Chaplygin LTB plus FLRW system must comply with the matching conditions (\ref{Darmois1})--(\ref{extraMC}) to a spatially flat FLRW background ($k_0=0$). Considering $r_{\rm b}=1$, the overdensity extends in the range $0\leq r\leq  1$ and the Chaplygin gas spatially flat FLRW background extends along $r>1$. 
\begin{figure}[htbp]
\begin{center}
\includegraphics[width=2in]{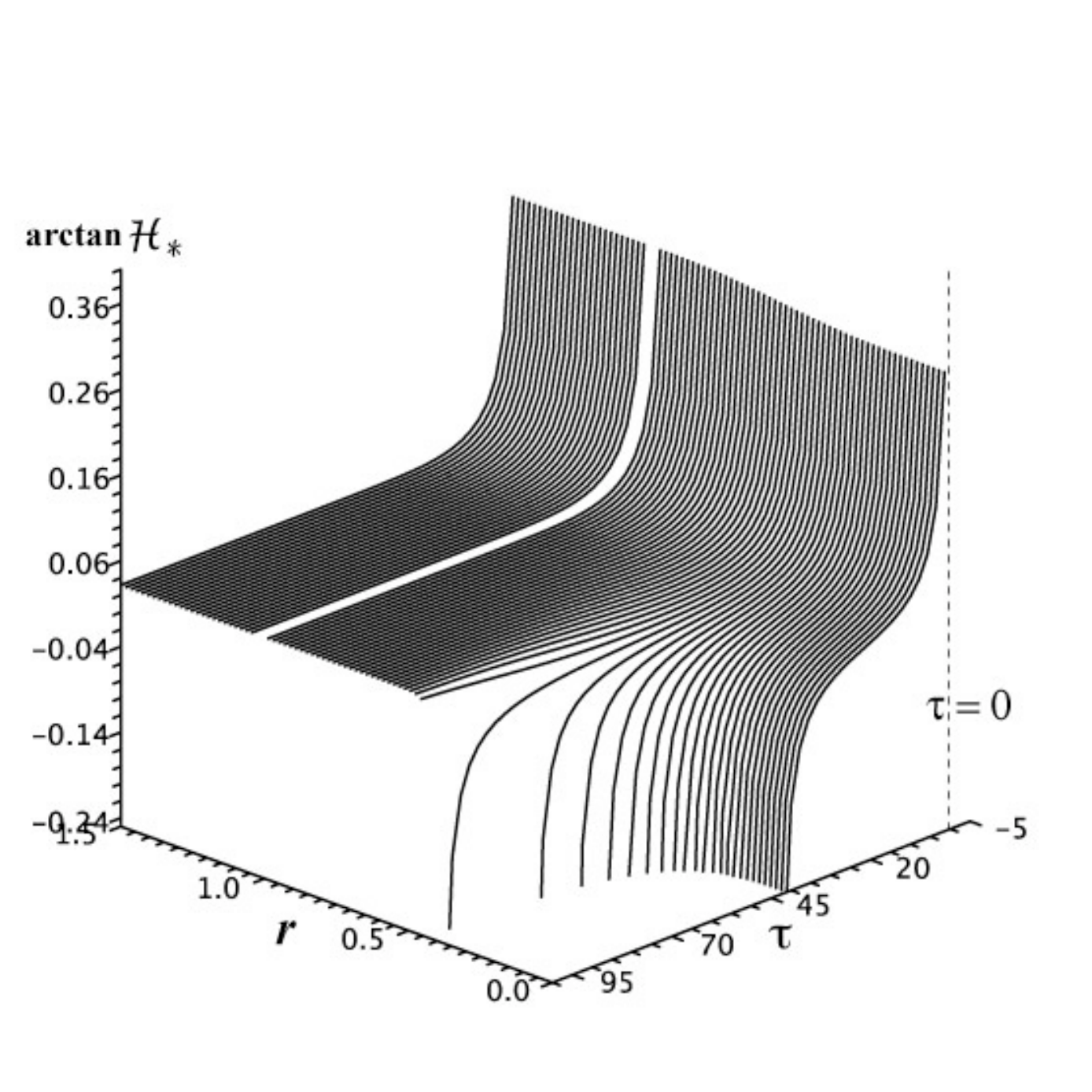}
\caption{{\bf Hubble expansion and collapse into a black hole.} The figure displays  the function
$\arctan \HH_*(\tau,r)$ for $\tau\geq 0$ corresponding to a Chaplygin gas overdensity ($0\leq r\leq
1$) smoothly matched to a Chaplygin gas universe at $r=r_{\rm b}=1$ (white strip). 
The dimension--less time $\tau$ is defined in (\ref{dimensionless2}) so that $\tau=0$ is the initial hypersurface. The function $\HH_*$
diverges at the initial big bang singularity. For inner layers it becomes negative and goes into
$\HH_*\to-\infty$ as these layers collapse into a black hole. For external layers blending into the
cosmic background $\HH_*\to\sqrt{\lambda}$, indicating an asymptotically de Sitter behavior.}
\label{fig1}
\end{center}
\end{figure}
\begin{figure}[htbp]
\begin{center}
\includegraphics[width=2in]{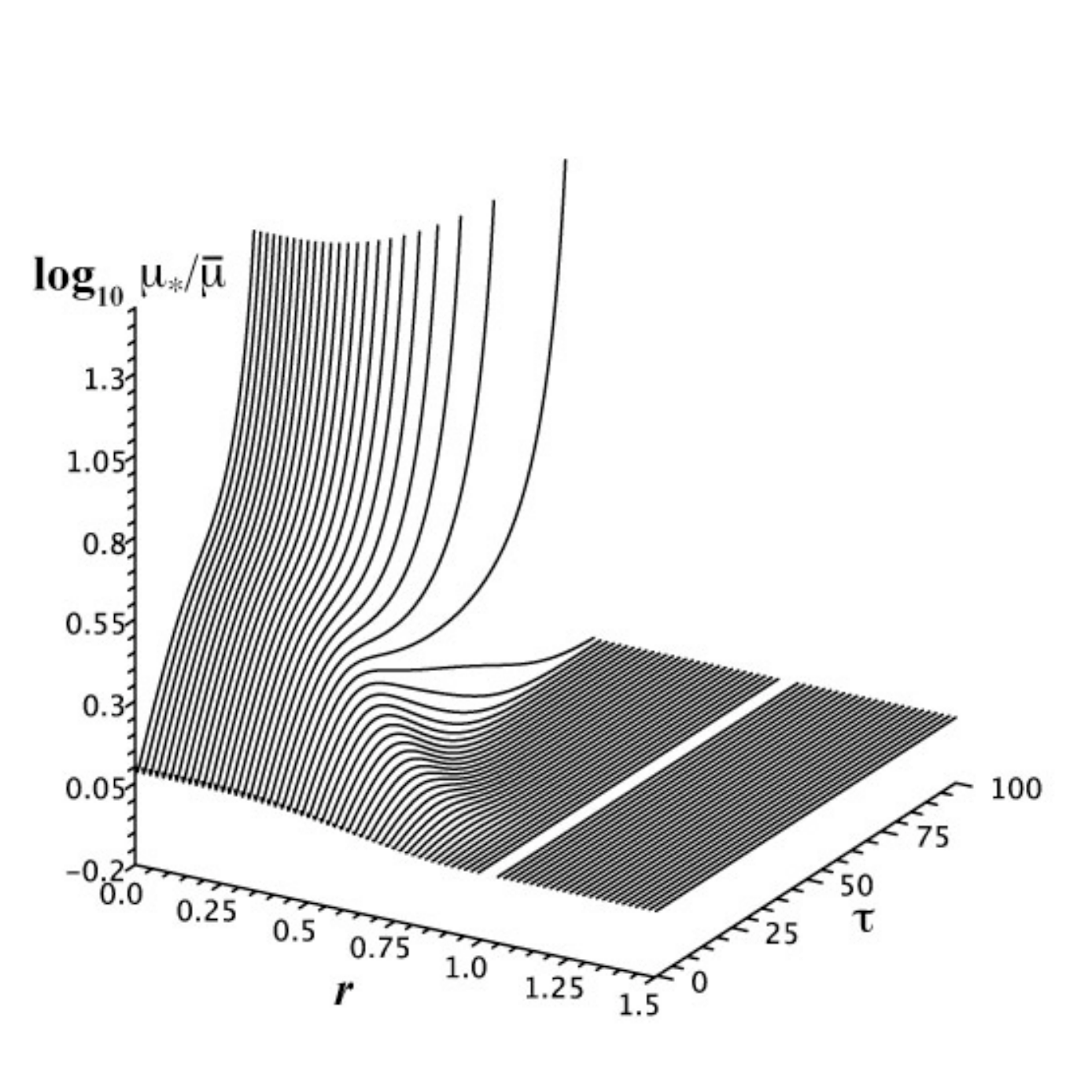}
\caption{{\bf Density contrast.} The figure depicts the ratio
$\mu_*(\tau,r)/\bar\mu(\tau)$ for a Chaplygin gas overdensity ($0\leq r\leq r_{\rm b}=1$) smoothly
matched to a Chaplygin gas FLRW universe ($r> 1$) with density $\bar \mu$. The matching interface
$r=r_{\rm b}=1$ is displayed as a white strip. The density contrast diverges as inner layers of the
overdensity collapse into a black hole, while external layers blend into the cosmic background as
$r\to 1$.}
\label{fig2}
\end{center}
\end{figure}
\begin{figure}[htbp]
\begin{center}
\includegraphics[width=2.5in]{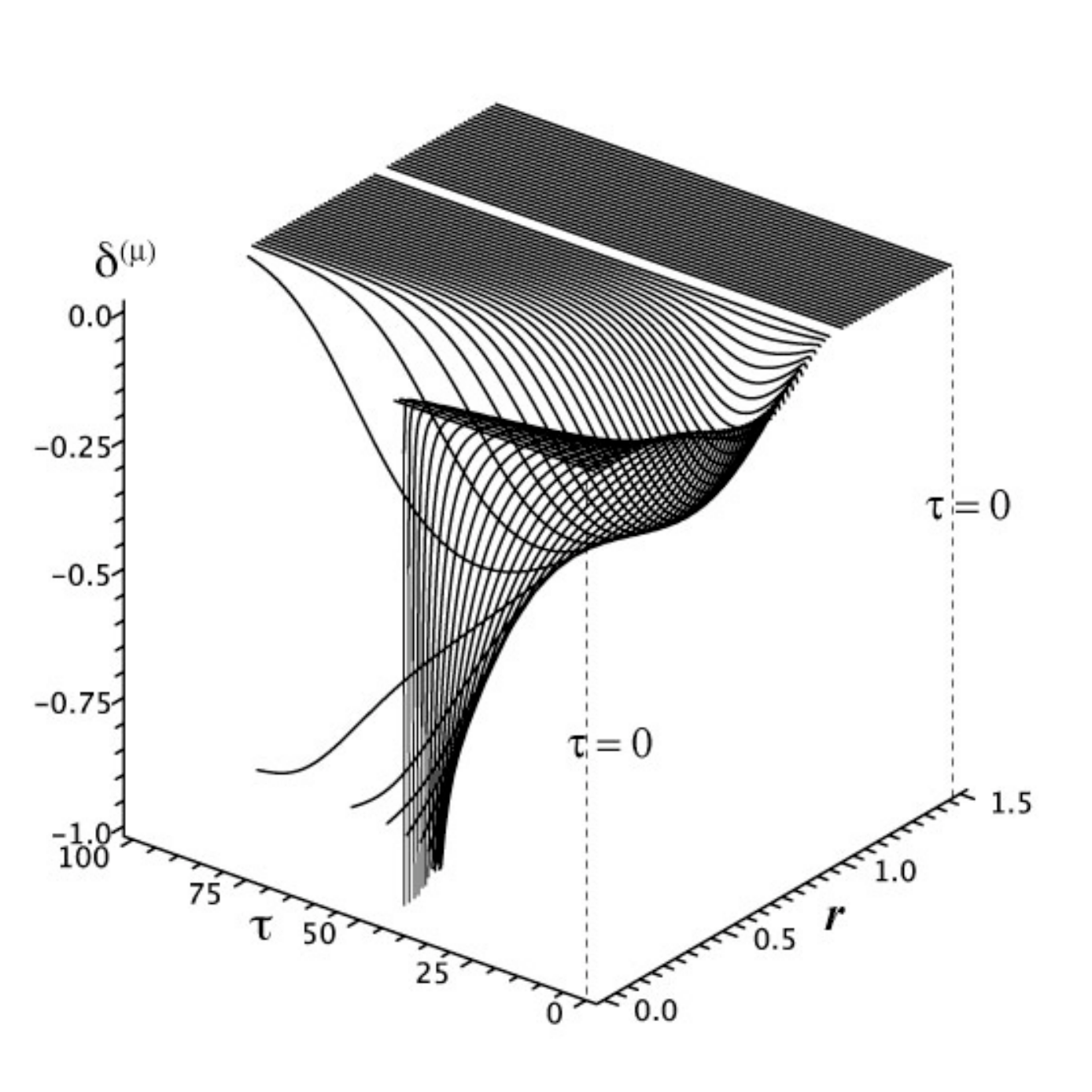}
\caption{{\bf Energy density fluctuations.} The figure displays the relative fluctuation, or
perturbations, $\Dm(\tau,r)$ for $\tau\geq 0$ corresponding to a Chaplygin overdensity smoothly
matched to a Chaplygin gas universe at $r=r_{\rm b}=1$ (white strip). This function is negative
(overdensity) and close to zero near the center and the matching interface (small radial gradient).
It is significantly different from zero in the areas where the radial gradient is large and $\Dm\to
-1$ as inner layers collapse into a black hole and at the initial singularity for all layers (not
shown).}
\label{fig3}
\end{center}
\end{figure}
\begin{figure}[htbp]
\begin{center}
\includegraphics[width=2in]{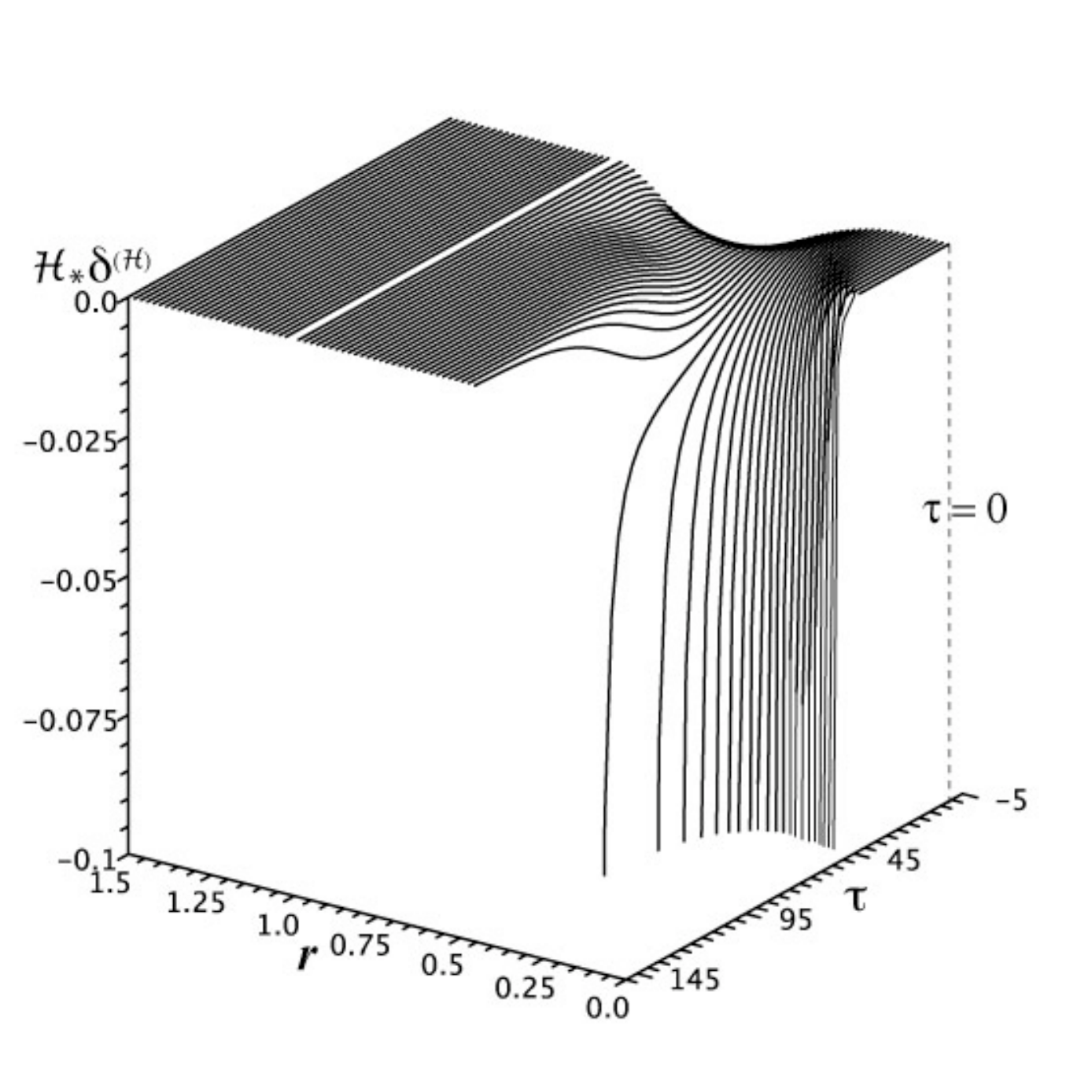}
\caption{{\bf Fluctuations of the Hubble scalar $\HH$.} Fluctuations of $\HH$ given by $\HH-\HH_*=\HH_*\Dh$. As in figure \ref{fig3}, it is close to zero near the center and the
matching interface and it is significantly different from zero in the areas where the radial
gradient is large. Notice that $\Delta \HH=-\Sigma$, where $\Sigma$ is the shear scalar function in
(\ref{Sigsc}) and (\ref{Sigma2}).}
\label{fig5}
\end{center}
\end{figure}

As expected on intuitive grounds,  inner layers in the inhomogeneous overdensity bounce and collapse, as the FLRW region keeps expanding. This is shown in the graphs of figures \ref{fig1} and
\ref{fig2}. The kinematic evolution of the overdensity follows from the behavior of the QL
Hubble scalar, $\HH_*=\Theta_*/3=\dot\ell/\ell$. As shown in figure \ref{fig1}, this function is
initially infinite at early times for all $r$ (overdensity and background), indicating an infinite
expansion rate from an initial big bang singularity at $\ell=0$. As $\tau$ increases and $\ell$
grows, the function $\HH_*$ monotonously deceases in the background and external layers of the
overdensity, but in inner layers we have $\HH_*\to 0$, which means a zero of $\dot\ell$ indicating
that inner comoving layers near $r=0$ reach maximal expansion. See equations
(\ref{ave_L})--(\ref{fried1}) and (\ref{Chap_EPav}). For later times (around
$\tau \sim 40-50$) we have $\HH_*\to -\infty$ for these layers, indicating an infinite contracting
rate associated with a re--collapsing singularity: curvature radius and scale factor, $R$ and
$\ell$, tend to zero, while $\mu_*$ diverges (from (\ref{Chap_Eav})). 

The diverging of $\mu_*$ for these layers is shown in figure \ref{fig2}, which displays the ratio
of the density $\mu_*$ in the inhomogeneous region to the cosmic background density $\bar\mu$
(the so--called ``density contrast'' in Newtonian astrophysical applications~\cite{tophats}). Notice that
$\mu_*/\bar\mu$ diverges at the same locus in the inner part of the overdensity where $\HH_*\to
-\infty$, while external layers of the overdensity smoothly blend to the background value:
$\mu_*/\bar\mu\to 1$.  

The evolution depicted by figures \ref{fig1} and \ref{fig2} is a clear signal of a ``big crunch''
singularity that forms a local black hole where $\HH_*\to-\infty$ and $\mu\to\infty$. On the other
hand, for the external layers of the overdensity $\HH_*$ remains always positive and we have
$\HH_*^2\to\lambda\sim 0.001$ and $\mu_*\to \lambda^2\sim 10^{-6}$ as $t\to\infty$ as they blend
into the cosmic background at $r=1$. This asymptotic behavior is consistent with the expanding
Chaplygin gas taking in this future asymptotic limit the form of a cosmological constant. 

Relative fluctuations of the energy density, $\Dm$, are shown for $\tau\geq 0$ in figure
\ref{fig3}. This function is identically zero for $r\geq 1$ since radial gradients are zero at the
cosmic background. It is negative but very close to zero in areas with negligible radial gradients
(near the center $r=0$ and near $r=1$). It is significantly different from zero for intermediate
values of $r$ where radial gradients are not negligible, while $\Dm\to -1$ as the internal layers
collapse into a local black hole (compare the locus of $\Dm\to-1$ with that of $\HH_*\to-\infty$
and $\mu_*\to\infty$). Since $\mu=\mu_*[1+\Dm]$, this behavior apparently implies that $\mu\to 0$
as $\mu_*$ diverges, which would be wholly unphysical. However, $\Dm\to -1$ does not imply $\mu\to 0$. As shown by (\ref{Dm_ana_ch}), $\Dm\to -1$ if $\Gamma\to\infty$. In fact, numerical numeric tests show that $\Gamma\to\infty$ as $\ell\to 0$ (big bang and re--collapse singularities), but $\ell^6\,\Gamma\to 0$ in these limits.  As $\ell\to 0$ both $\mu$ and $\mu_*$ diverge because the product $\mu_*[1+\Dm]$ diverges even if $1+\Dm\to 0$.

It is important to mention that pressure anisotropy can be directly related to $\Dm$ by means of
(\ref{PP2}) and (\ref{fluc_Chap}), leading to the ratio
\begin{equation}\frac{\PP}{p} = \frac{-\Dm}{2(1-\Dm)},\label{anisp}\end{equation}
hence, pressure is nearly isotropic when radial gradients are negligible and reaches its maximal
anisotropic ratio\, $\PP/p=1/4$ \, as \, $\Dm\to-1$ \, in the singularities, which can then be
characterized as anisotropic (as well as non--simultaneous).     

The relative fluctuations of the Hubble scalar $\Dh=(\HH-\HH_*)/\HH_*$ will necessarily diverge for
inner layers at their maximal expansion ($\HH_*=0$). Hence, we plotted instead in figure \ref{fig5}
the fluctuations $\HH-\HH_*=\HH_*\Dh=-\Sigma$. As shown by this figure, $\HH-\HH_*=0$ for $r\geq 1$
(fluctuations vanish at the cosmic background) and is negative but small for most of the range of
$r$ and $\tau$ (``clumps'' of $\HH$), but as $\HH_*$ (see figure \ref{fig1}) $\HH-\HH_*\to -\infty$
at the collapsing singularity.

\subsection{Void formation in an interactive CDM-DE mixture}\label{Imixture}

A phenomenological interactive CDM--DE mixture in a FLRW background  was examined in the three
articles of reference \cite{intmix2} and reviewed more recently in \cite{intmix_new}. 
This model assumes that the CDM component is dust and the DE component is a fluid satisfying 
a simple linear EOS of the type (\ref{BEOSlin}). It corresponds
to the particular case of (\ref{newnames}) given by
\ba\varphi_* &=& \delta_\varphi=0, \qquad w_\rho=0,\nonumber\\ 
\pi_* &=& w_0\,q_*,\qquad \delta^{(\pi)}=\Dq,\qquad w_q=w_0,\label{IM_pars}\ea
where $w_0$ is a constant in the range $-1<w_0<-1/3$. Notice that this type of  linear EOS is
also satisfied by local variables: $\pi=w_0 q$. The interaction is assumed to be proportional to
the conservation term of the CDM dust component: 
\begin{equation}J_* =3\,\epsilon\,\rho_*\,\HH_*,\label{int_term1}\end{equation}
where $\epsilon$ is an adjustable constant parameter (to compare with the notation of equation (19) 
of \cite{intmix_new} we have $\epsilon = -\alpha$). For this functional  dependence of $J_*$
we get
\begin{equation}\delta^{(J)}=\frac{[\ln J_*]'}{3R'/R}=\frac{[\ln \rho_*+\ln\HH_*]'}{3R'/R}  =
\Drho+\Dh.\label{int_term2}\end{equation}
Using (\ref{deltas_mixt}), it is straightforward to verify that (\ref{int_term1})  and
(\ref{int_term2}) are consistent with $J=3\,\epsilon\rho\HH$ in the local conservation law
(\ref{local_cons}). The evolution equations are the following particular case of (\ref{eveqs_mixt}):
\begin{widetext}\bse\label{eveqs_mixt2}\ba \dot\HH_*&=& -\HH_*^2
-\frac{\kappa}{6}[\rho_*+(1+3w_0)q_*],\\
\dot\rho_* &=& -3(1-\epsilon)\,\rho_*\,\HH_*,\label{rhoavt}\\
  \dot q_* &=& -3\,[(1+w_0)\,q_*+\epsilon\rho_*]\,\HH_*,\label{qavt}\\
   \dDrho &=& -3\,\HH_*\,\Dh\,[1-\epsilon+\Drho],\\
    \dDq &=& -3\HH_*\,\left[(1+w_0+\Dq)\,\Dh  -\epsilon\frac{\rho_*}{q_*}(\Dq-\Drho-\Dh)\right],\\
 \dDh &=& -\HH_*\Dh(1+\Dh)+\frac{\kappa}{6\HH_*}
\left[\rho_*(\Dh-\Drho)+(1+3w_0) q_* (\Dh-\Dq)\right],
\ea\ese\end{widetext}
which is a completely determined system depending on two constant free parameters: $w_0$ and
$\epsilon$.

\begin{figure}[htbp]
\begin{center}
\includegraphics[width=2.5in]{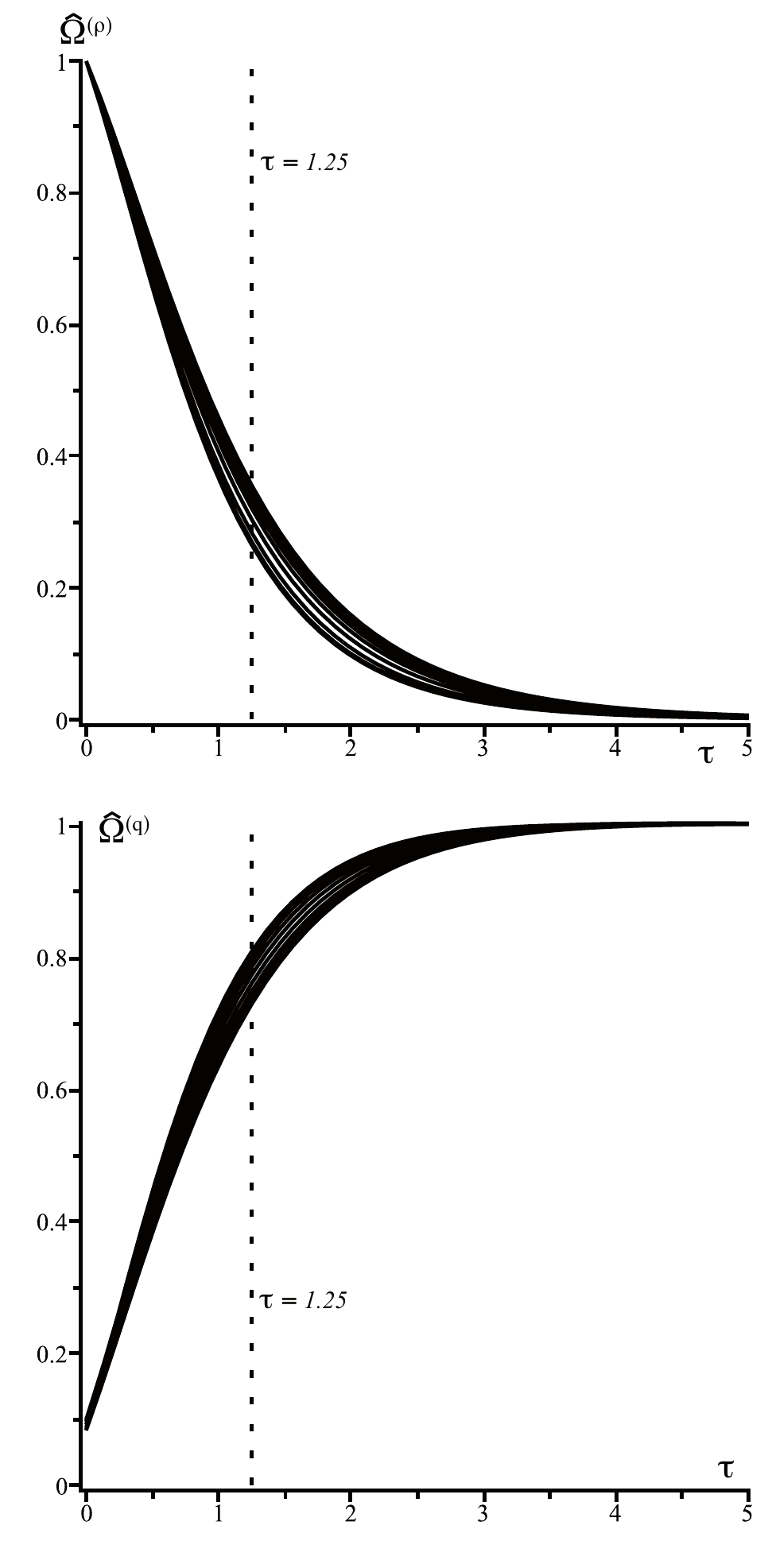}
\caption{{\bf The Omega parameter for CDM and DE.} The figure displays the time evolution for the
full radial range of the parameters $\hat\Omega^{(\rho)}$ (top panel) and $\hat\Omega^{(q)}$
(bottom panel) defined by (\ref{hatOm}). We can roughly identify the present day era as
$\tau=1.25$.}
\label{fig6}
\end{center}
\end{figure} 
\begin{figure}[htbp]
\begin{center}
\includegraphics[width=2.5in]{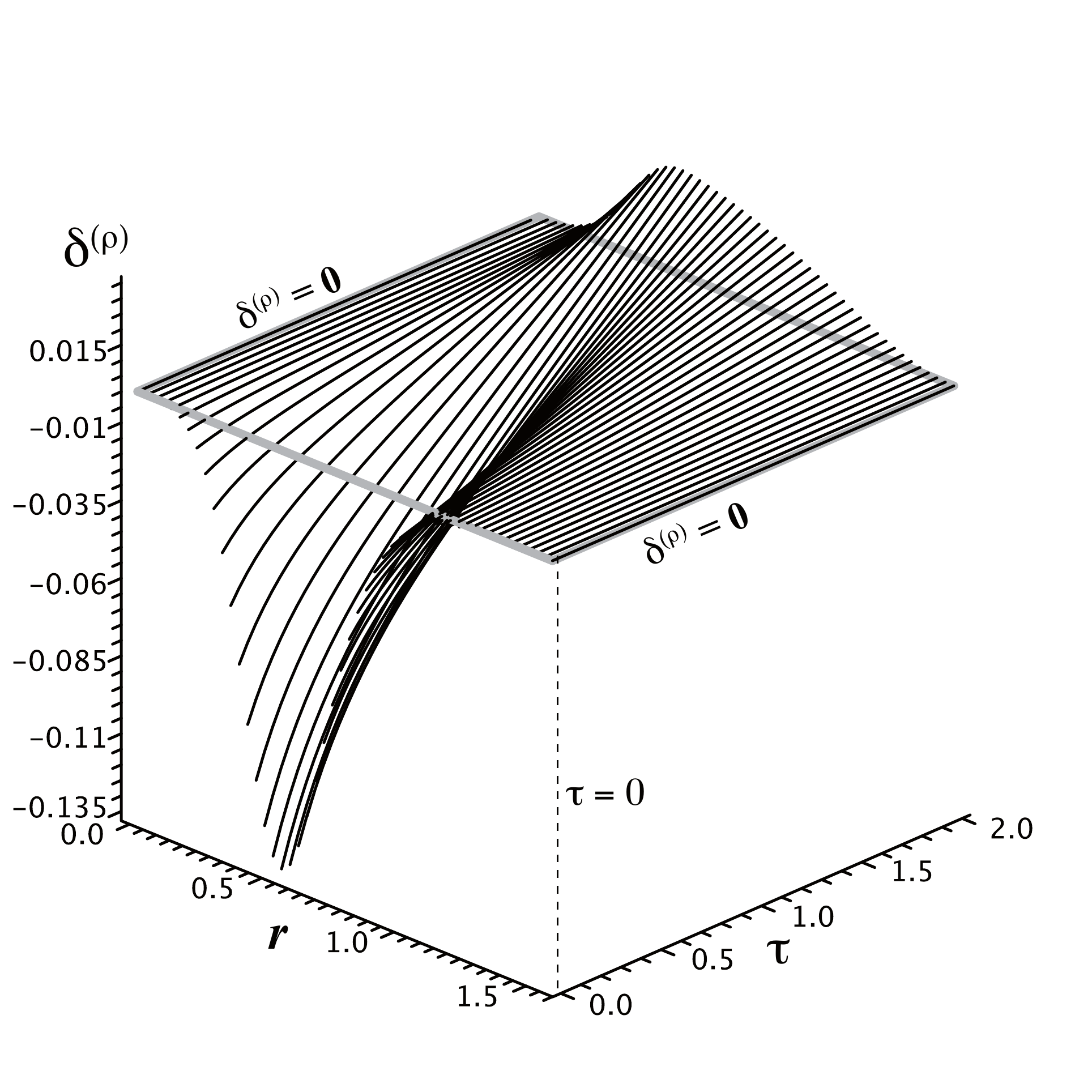}
\caption{{\bf Relative fluctuations of CDM density.} The figure displays the relative fluctuation
of CDM, $\Drho$. This function passes from negative to positive values for all $r$, indicating
density clumps evolving towards density voids, as shown in figure 9.}
\label{fig7}
\end{center}
\end{figure}
\begin{figure}[htbp]
\begin{center}
\includegraphics[width=2.5in]{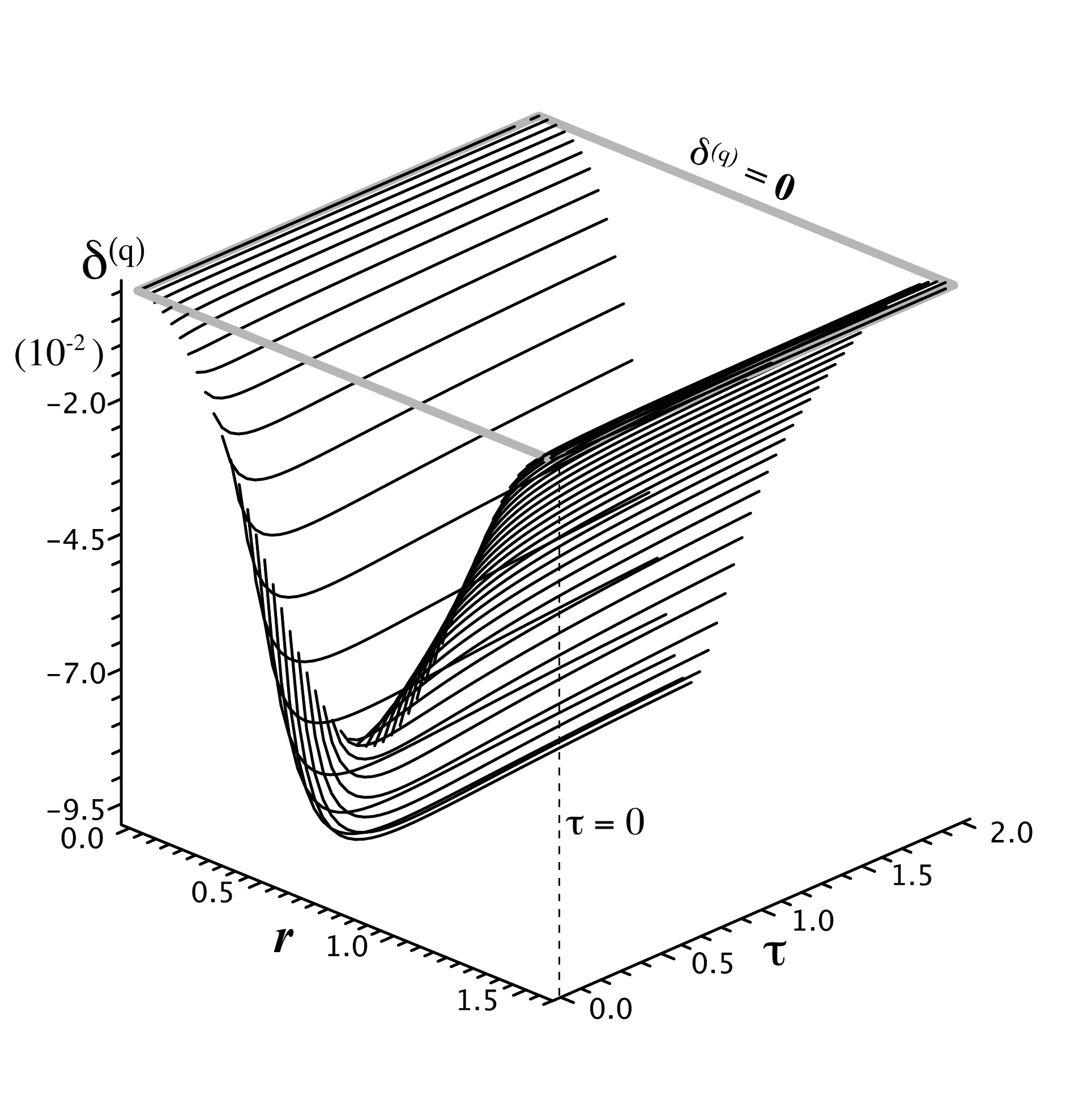}
\caption{{\bf Relative fluctuations of DE density.} The figure displays the  function $\Dq$. This
function takes small negative values ($\sim 10^{-2}$) for all layers, indicating low depth density
clumps evolving towards homogeneity. Since pressure anisotropy is the fluctuation of DE pressure,
it is proportional to $\Dq$ and so we have $\PP/p\sim 10^{-2}$.}
\label{fig8}
\end{center}
\end{figure} 
\begin{figure}[htbp]
\begin{center}
\includegraphics[width=3in]{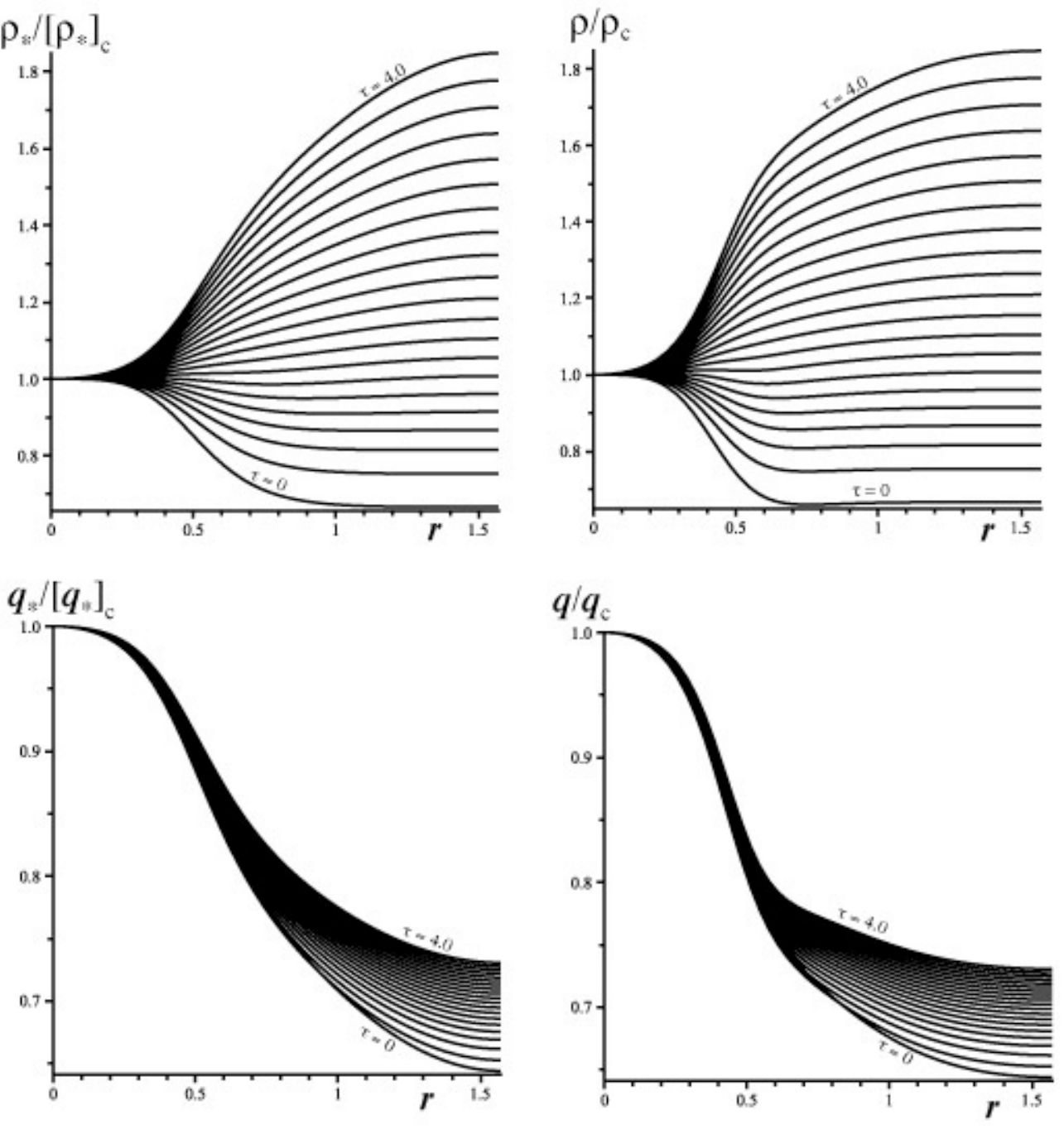}
\caption{{\bf Radial profiles of CDM and DE energy densities.} The panels display the QL and
local densities for CDM (top two panels) and DE (bottom panels), normalized by their central
values, as functions of $r$ for the following constant values running from $\tau=0$ to $\tau=4.0$
in intervals of $0.2$, from bottom to top. The curves show how local and QL variables
exhibit the same qualitative behavior. The figure confirms that an initial CDM density clump
evolves towards a deep density void, while DE shallow clumps slowly tend to
homogeneity.}\label{fig9}
\end{center}\end{figure}

A frequent motivation to consider interaction between CDM and DE in the numerous  works in the
literature~\cite{intmix1,intmix2} has been to address the so--called ``coincidence problem''. In
the FLRW solutions of \cite{intmix2}, the energy densities of CDM and DE are found as a function of
the scale factor, and the authors adjust $\epsilon$ in order to obtain a ratio of these energies
that displays an asymptotically slowly varying behavior. Since we are identifying FLRW energy
densities with the background variables  $\rho_*$ and $q_*$, this procedure involves verifying the asymptotic behavior of
the ratio $\rho_*/q_*$, both expressed as functions of the scale factor $\ell$. Inserting
$\mu=\rho+q $ and $p=w_0 q$ into (\ref{fried11}), and replacing $w_0$ with $w_0=-|w_0|$, we obtain
\bse\label{MQavIM}\ba \rho_* &=& \frac{[\rho_*]_i}{\ell^{3(1-\epsilon)}},\label{MavIM}\\ 
        q_* &=&
\frac{(|w_0|-\epsilon)[q_*]_i-\epsilon[\rho_*]_i(1-\ell^{-3(|w_0|-\epsilon)})}{(|w_0|-\epsilon) 
\,\ell^{3(1-|w_0|)}},\nonumber\\\label{QavIM}\ea\ese
so that for $|w_0|>\epsilon$ and $\ell\gg 1$ we have
\begin{equation}\frac{\rho_*}{q_*}\sim \,\ell^{-3(|w_0|-\epsilon)}.\end{equation}
Hence, for $\epsilon>0$ smaller than but sufficiently close to $|w_0|$,  the ratio $\rho_*/q_*$
would decay very slowly. However, we will follow here a different motivation.    

The conservation equation (\ref{local_cons}) in its form (\ref{rhoavt}) can also be understood as
describing an interaction in which total CDM particle numbers increase ($\epsilon>0$) or decrease
($\epsilon<0$)~\cite{intmix_new,lima}. These numbers would be conserved if $\epsilon=0$, leading to a non--interactive
mixture. If we  consider the case with spatially flat hypersurfaces $\T$ (see Appendix B), which follows as the
particular case $[k_*]_i=K= 0$ for all $r$ in (\ref{LTB}) and (\ref{LTB2}), the proper spatial volume associated
to these spatially flat metrics leads then exactly to the same integral used in the definition of QL variables
(\ref{QLfunc}). Thus, the total number of CDM particles inside a comoving sphere (centered at $r=0$) in any given
hypersurface $\T$ is 
\begin{equation} N = 4\pi\int_0^r{n\,R^2R'dx}=\frac{4\pi}{3}n_* R^3,\label{Ndef}
\end{equation}
where $n=\rho/m_0$ is the particle number density for CDM particles of rest mass $m_0$. With the
help of (\ref{ave_L}) and (\ref{rhoavt}), and bearing in mind that $n_*=\rho_*/m_0$, equation (\ref{Ndef}) leads to
the following particle number scaling law: 
\begin{equation} N = N_i\,\ell^{3\epsilon}.\label{Ncdm}\end{equation}
where $N_i(r)$ is an initial particle number distribution. Since at earlier stages the universe must have
been dominated by CDM while presently DE dominates, it is reasonable to consider a scenario with
CDM particle numbers decaying and so (\ref{MQavIM}) and (\ref{Ncdm}) suggest considering
$\epsilon=-|\epsilon|<0$, but with $|\epsilon|<|w|$. Instead of the ``coincidence problem'' we
consider how the dynamics of the mixture is affected by decaying numbers of CDM particles in a late time universe. 

For the numerical examination of the evolution equations (\ref{eveqs_mixt2}), we consider for the
spacially flat case $[k_*]_i=0$ the coordinate choice (\ref{Yi_gen}) with $f=\tan r$,
 and the parameter values:\, $w=-0.9$ that is close to a cosmological
constant, and $\epsilon=-0.1$, so that CDM particles decay. Appropriate initial value functions
that comply with an evolution free from shell crossings (satisfying with condition (\ref{no_xcr}))
are given in Appendix D. 

The central and asymptotic values given by the initial conditions in (\ref{ICIM}) (Appendix D)
are adequate for density clumps in a matter
dominated era. Hence, we can take $h_s^{-1}$ as an early universe Hubble scale factor. In order to
verify that the evolution of the mixture yields CDM and DE energy densities that are minimally
compatible with observations, we define the following Omega parameters:
\bse\label{hatOm}\ba \hat\Omega^{(\rho)} \equiv \frac{\kappa\,\rho_*}{3\,\HH_*^2},\\
        \hat\Omega^{(q)} \equiv \frac{\kappa\,q_*}{3\,\HH_*^2}, \ea\ese
whose time evolution is plotted in figure \ref{fig6} for the range $0\leq r<\pi/2$. As shown by
this figure, CDM is initially dominant with $\hat\Omega^{(\rho)}\approx 1$ and
$\hat\Omega^{(q)}\approx 0$ at the initial time $\tau=0$, while as $\tau$ proceeds
$\hat\Omega^{(\rho)}\to 0$ and $\hat\Omega^{(q)}\to 1$. If instead of (\ref{hatOm}), we use Omega
parameters defined with local quantities, $\rho,\,q$ and $\HH$, we obtain almost identical curves.
From figure \ref{fig6} we can roughly identify present cosmic era values around $\tau=1.25$. 

In order to examine the density profiles of CDM and DE it is useful to recall from (\ref{prop1})
that:
\bse\label{Deltadefs}\ba \Drho &=& \frac{1}{\rho_* R^3}\,\int_0^r{\rho'\,R^3\,dx},
\label{Drhodef}\\
\Dq &=& \frac{1}{q_* R^3}\,\int_0^r{q'\,R^3\,dx},\label{Dqdef}\ea\ese
so that (for positive $\rho_*$ and $q_*$) the sign of the relative fluctuations determine whether $\rho$ and $q$ present radial
profiles of over--densities or voids in a domain of $r$ around a symmetry center. In particular, it
is possible to provide initial conditions in which an initial over--density evolves into a void.
This situation has been reported in dust sources \cite{suss08,mustapha}, and in particular, initial
conditions (\ref{ICIM}) yield such an evolution for the CDM densities $\rho_*$ and $\rho$. 

The relative fluctuation $\Drho$ is depicted by figure \ref{fig7}, showing how for all layers this
function passes from negative to positive as $\tau$ evolves. Hence, from (\ref{Drhodef}), we have
an initial CDM clump evolving into a configuration with radial profiles of void. Notice that near
the center $r=0$ and the asymptotic range $r\to \pi/2$ we have $\Drho$ close to zero, nevertheless,
it passes from negative to positive values for all layers. This ``clump into void'' evolution is
corroborated in the top panels of figure \ref{fig9}, which displays for various values of $\tau$
the radial profiles of $\rho$ and $\rho_*$, normalized to their central values. As revealed by this
figure, it is striking to see how an initial CDM density clump profile evolves into a clear and
deep void. The values of $\tau$ start at $\tau=0$ and go up to $\tau=0$ in steps of $0.2$, so if we
take $\tau\approx 1.25$ as indicative of our present era (figure \ref{fig6}), then the top panels
of figure \ref{fig9} predict that the CDM radial profile in our era is a very shallow clump, or in
other words, that we would be in the middle of the transition from clumps into voids. 

On the other hand, as shown by figure \ref{fig8}, the relative fluctuation of DE, $\Dq$, is
negative and small ($\sim 10^{-2}$) for all layers and for all $\tau$. Hence, the radial DE profile
is that of shallow clumps that slowly tend to homogeneize. This evolution is confirmed by plotting
in the lower panels of figure \ref{fig9} the radial profiles of DE energy density $q$,\,$q_*$
(normalized by their central values) for various values of constant $\tau$.  From (\ref{PP2}),
(\ref{deltas_mixt}) and (\ref{IM_pars}) the ratio of anisotropic to isotropic pressures is
$2\PP/\pi=\Dq/(1+\Dq)\approx \Dq\sim 10^{-2}$, and so this anisotropy is very small.  

While the `clump becoming void' evolution that we have found (see figures \ref{fig7} and top panels
of figure \ref{fig9}) is still present even when $\epsilon=0$ (no interaction, conserved $N$), the
effect of a larger (more negative) $\epsilon$ is to increase the depth of the resulting voids and
the speed at which they occur. This is consistent, because as CDM particle numbers decay faster one
would expect the voids to be deeper and to arise faster. At the same time, the DE energy density profile
becomes homogeneous at a faster rate. However, if we take $\epsilon>0$, so that CDM particles are
created, then we get the opposite effect: the DE profiles pass from clumps to voids and CDM has the
profile of clumps. We have used here very idealized parameters and do not claim that the model
mixture is ``realistic'', but by following a more careful and comprehensive approach we could
consider this type of models as describing mechanisms to explain the relation between generation of
voids and an interaction involving the annihilation of CDM particles into DE. We will examine this
matter in a separate article. 

\section{Conclusion}

We have presented a formalism, based on quasi--local (QL) variables, describing a large class of
spherically symmetric models (LTB spacetimes) as non--linear, gauge invariant and covariant
perturbations of a FLRW formal background. The integration of the resulting evolution equations 
require numerical work, but this task can be handled with relatively simple numerical techniques. 
A summary of how this formalism was motivated, introduced, developed and applied is furnished in the Introduction (section II). 

Since LTB spacetimes are compatible with a wide variety of ``equations of state'' (EOS) and physical assumptions, these models and the formalism developed for them are useful theoretical tools to examine a large amount of cosmological sources of interest that have been studied only in a FLRW context (or linear perturbations)~\cite{review,Lambda,chaplygin1,chaplygin2,intmix1,intmix2,intmix_new,EOS1,EOS2}. While the inhomogeneity of LTB spacetimes is certainly idealized, the resulting models are non--trivial and still exhibit non--linear behavior, and can be useful to examine important phenomena (gravitational collapse, void formation) that cannot be studied with FLRW models or their linear perturbations. 

As we argued in section \ref{EOS}, the EOS that determines the evolution equations is given between the QL energy density and pressure, so that local density and pressure and pressure are related by means of correlation terms that could contain non--local information that could be important for self--gravitating sources. Since the fundamental nature of dark matter and dark energy is not known, we cannot rule the possibility of non--local effects playing an important role when we examine these sources under inhomogeneous conditions, at least in a scale smaller than 300 Mpc. We believe that LTB models can be useful to examine these and other non--linear effects, not only for dark matter or dark energy, but also in theoretical proposals that seek to explain cosmic acceleration without dark energy~\cite{Inh_rew1,Inh_rew2,InhObs1,InhObs2,ave1,ave2,ave3,theo1,theo2,ltbave,condsBR,wiltshire}. As shown elsewhere~\cite{condsBR,QLvars}, the formalism presented here can be applied to understand the issue of ``back--reaction'' that appears in various averaging formalisms~\cite{ave1,ave2,ave3} aiming to explore this type of alternative explanations.  

An important result of this article is the smooth and fully relativistic generalization of the Newtonian ``top hat'' models (section \ref{tophats}), which are useful toy models for structure formation and are widely used in Astrophysical literature. One of the numerical examples of application of the formalism that we provided (section \ref{chaplygin}) is a Chaplygin gas ``top hat'' model that describes the formation of a local black hole smoothly embedded in an expanding Chaplygin gas FLRW universe. 

The second numerical application example is that of an interactive mixture of cold dark matter (CDM) and dark energy. By interpreting the interaction term in terms of particle creation, we obtained a model in which initial clump (or overdensity) radial profiles of CDM evolve into void profiles as CDM decays into dark energy. While this model is not ``realistic'', it serves to illustrate how LTB spacetimes can illustrate important non--linear features of dark energy/matter sources that cannot be studied within a homogeneous context (nor with linear perturbations). As shown in \cite{intmix2,intmix_new}, mixtures of this type can be studied by means of a dynamical systems approach. This type of approach is compatible with the evolution equations of LTB spacetimes (autonomous equations), and so a generalization can be readily made of these dynamical system studies in a FLRW context. In fact, a dynamical systems study of LTB dust solutions has been achieved using the same QL variables~\cite{suss08}. The extension of this work for LTB spacetimes with nonzero pressure is being currently undertaken.

\section*{Appendix A: Initial conditions.} 

One possibility for setting up the initial value functions for the system (\ref{eveqs_ql}) for an EOS (\ref{BEOS}) is simply to select functional forms for
\begin{equation}[\mu_*]_i,\,[\HH_*]_i,\qquad [p_*]_i=[p_*]_i([\mu_*]_i),\label{initc1}
\end{equation}
(subscript ${}_i$ denotes evaluation at $t_i=0$, or $\tau=0$) and then, bearing in mind that $\Gamma_i=1$, and using (\ref{rad_grads2}) to obtain
\begin{equation}\Dim=\frac{r}{3}\,\frac{[\mu_*]_i'}{[\mu_*]_i}\qquad \Dih=\frac{r}{3}\,\frac{[\HH_*]_i'}{[\HH_*]_i}.\label{initc2}\end{equation}
However, a more intuitive choice involves considering as an initial value function the scalar  curvature $[k_*]_i=[\KK_*]_i/h_s^2$, since there is a connection between the kinematic evolution of the scale factor $\ell$ (or $L$) and the sign of this curvature. This follows from the Friedman equation (\ref{fried12}), so that $[k_*]_i>0$ is a necessary condition for
$0=\dot\ell=\HH_*$, indicating a  ``bounce'' of $\ell(t,r)$. On the other hand, $[k_*]_i\leq 0$ is a sufficient condition for $\ell$ being a monotonous function of $t$. 

The fact that the sign of $[k_*]_i$ determines the kinematic behavior of $\ell$ is analogous to the relation of the scale factor $a$ to the constant ``curvature index'' $k_0$ in FLRW spacetimes. The difference is that in FLRW spacetimes $k_0$ is a constant and thus all comoving layers in a given spacetime have the same
kinematic evolution, while in LTB spacetimes $[k_*]_i=[k_*]_i(r)$. Therefore, if this function changes sign in its domain of $r$ various comoving layers in the same LTB spacetime can have different kinematic evolution (as in the example of the Chaplygin gas in section \ref{chaplygin}).

In order to set initial conditions in terms of $\KK,\,\KK_*$ and $\Dk$, we  note that these curvature scalars can be obtained $\mu_*,\,\HH_*,\,\Dm,\,\Dh$ from the Friedman equation (\ref{cHam2}) and (\ref{rad_grads}):
\bse\label{KdK}\ba \KK_* &=&  \frac{\kappa}{3}\mu_* -\HH_*^2,\label{KmH}\\
\KK_*\,\Dk &=& \frac{\kappa}{3}\mu_*\Dm-2\HH_*^2\Dh,\label{DKDmDH}\ea\ese
so that we can always eliminate either one of $\mu_*,\,\HH_*$ or $\Dm,\,\Dh$ in terms of $\KK$ and $\Dk$. Thus, an alternative  set of initial conditions follows by selecting      
\begin{equation}[\mu_*]_i,\,[\KK_*]_i,\qquad [p_*]_i=[p_*]_i([\mu_*]_i),\label{initc3}
\end{equation}
and then $\Dim$ as in (\ref{initc2}), while $[\HH_*]_i$ and $\Dih$ follow from (\ref{rad_grads2}) and (\ref{KdK}) as
\ba[\HH_*]_i &=& \left\{\frac{\kappa}{3}[\mu_*]_i-[\KK_*]_i\right\}^{1/2},\nonumber\\  \Dih &=&
\frac{r}{6}\,\frac{(\kappa/3)\,[\mu_*]_i'-[\KK_*]_i'}{(\kappa/3)\,[\mu_*]_i-[\KK_*]_i}.\label{initc4}
\ea
We have used initial conditions given by (\ref{initc3}) and (\ref{initc4}) in the numeric examples of section \ref{applications}.  

\subsection*{Appendix B: Special initial conditions: the zero curvature case.}

It is evident from equations (\ref{ave_K}) and (\ref{KdK}) that the following  choice of initial conditions:
\begin{equation}[\KK_*]_i = \Dik=0, \quad \Rightarrow\quad \KK_* =  \Dk =
0,\label{zeroK_ic}\end{equation}
leads to a simplified and important particular case that can be identified as LTB spacetimes with zero spatial curvature. Conditions (\ref{zeroK_ic}) lead to the constraints:
\begin{equation} \HH_*^2=\frac{\kappa}{3}\mu_*,\qquad \Dh=\frac{\Dm}{2},
\label{zeroK_cons}\end{equation}
which allows us to eliminate $\HH_*$ and $\Dh$  in terms of $\mu_*$ and $\Dm$. The evolution equations (\ref{eveqs_ql}) then simplify considerably and become the following system of two equations:
\bse\label{eveqs_zeroK}\ba \dot\mu_* &=& -\sqrt{3\kappa}\,\mu_*^{3/2}\,
\left[1+w\right],\\ \dDm &=&
\frac{\sqrt{3\kappa}}{2}\,\mu_*^{1/2}\,\left[(\Dm-2\Dp)\,w-\Dm\, (1+\Dm)\right].\nonumber\\\ea\ese
where $w=p_*/\mu_*$ and there is no restriction on the EOS. These equations can  be integrated analytically for many choices of EOS (see \cite{intmix3,hydro}), but, in general, they also need to be integrated numerically. Initial conditions (\ref{zeroK_ic}) were used in the mixture example of section \ref{Imixture}.

\subsection*{Appendix C: Initial conditions for the Chaplygin gas.}

Dimensionless forms for the initial value functions $[\mu_*]_i,\,[p_*]_i,\, [\HH_*]_i$ for the Chaplygin gas ``top hat'' model of section \ref{chaplygin} follow from applying the Chaplygin gas EOS to (\ref{rad_grads2}) and (\ref{initc4}), and using the prescription (\ref{dimensionless1}) and the dimensionless parameters $\lambda$ and $m_i$ defined in (\ref{Chap_pars}). The resulting initial
value functions are:
\bse\label{CH_IVF}\ba 
  \tilde\Omega_i^{(\mu)} &\equiv& \frac{\kappa[\mu_*]_i}{3h_s^2} = 
[\,m_i^2+\lambda^2\,]^{1/2},\label{Chap_pars3}\\
  \tilde\Omega_i^{(p)} &\equiv& \frac{\kappa [p_*]_i}{3h_s^2} = 
-\frac{\lambda^2}{[\,m_i^2+\lambda^2\,]^{1/2}},\\
   \frac{[\HH_*]_i^2}{h_s^2} &=& \tilde\Omega_i^{(\mu)}-[k_*]_i,\\
 \Dim &=& \frac{r}{3}\frac{[\tilde\Omega_i^{(\mu)}]'}{\tilde\Omega_i^{(\mu)}},\\
\Dih &=& \frac{r}{3}\frac{[\HH_*]_i'}{[\HH_*]_i}=
\frac{r}{6}\frac{[\tilde\Omega_i^{(\mu)}]'-[k_*]_i'}{\tilde\Omega_i^{(\mu)}-[k_*]_i},
\ea\ese
where $k_i=[\KK_*]_i/h_s^2$. 

The functions in (\ref{CH_IVF}) must lead to solutions of the evolution equations complying with the matching and regularity conditions (\ref{Darmois1})--(\ref{extraMC}) and with the condition to avoid shell crossings (\ref{no_xcr}). Initial value functions satisfying the above mentioned requirements
follow by taking $\lambda=0.001$ and assuming that $\tilde\Omega_i^{(\mu)}$ and $[k_*]_i$ are given by the simple polynomial ansatz
\begin{equation} Z = \left\{ \begin{array}{l} Z_c  - x^{4} \left( {15-24x +10 x^2} 
\right)\left( {Z_c  - Z_b } \right),\quad 0 \le x \le 1 \\  Z_b ,\hskip 5.5cm x \ge 1 \\ 
\end{array} \right.\label{Zdef}\end{equation}
where $x\equiv r/r_b$, so that $Z_c=Z(0)$ and $Z_b=Z(1)$. Central and background values are given by 
\bse\label{ICchap}\ba \tilde\Omega_i^{(\mu)}(0) &=& 1.1,\qquad
\tilde\Omega_i^{(\mu)}(1)=1.0,\\
\left[k_*\right]_i(0) &=& 0.1,\qquad \left[k_*\right]_i(1)=0,\ea\ese
corresponding to an initial overdensity with small density contrast and with small positive curvature matched to a spatially flat background. 

\subsection*{Appendix D: Initial conditions for the interactive CDM-DE mixture.}

In the example of the interactive DM--DE mixture of section \ref{Imixture}, we chose initial value functions for (\ref{eveqs_mixt2}) by considering (instead of (\ref{Yi1})) the following choice of radial coordinate:
\begin{equation} R_i = h_s^{-1}\,\tan\,r,\label{Yi2}\end{equation}
so that (\ref{Gdef}) and (\ref{rad_grads2}) become
\bse\ba \Gamma = 1+\frac{\ell'/\ell}{R_i'/R_i}=1+\cos r\,\sin r\,\frac{\ell'}{\ell},\label{Gdef2}\\
\Da = \frac{1}{3\Gamma}\,\frac{A_*'/A_*}{R_i'/R_i}=\frac{\cos r\,\sin
r}{3\Gamma}\,\frac{A_*'}{A_*}.\label{newdelta2}\ea\ese
This choice allows us to use a finite coordinate range $0\leq r<\pi/2$ to examine the full asymptotic range along the $\T$ associated with $R_i\to\infty$. 

Since we considered also the spatially flat case $k_i=[\KK_*]_i=0$, then (\ref{initc4}) becomes
\ba [\HH_*]_i &=& \left[\frac{\kappa}{3}\left([\mu_*]_i-[q_*]_i\right)\right]^{1/2},\nonumber\\  \Dih &=&
\frac{\cos r\,\sin
r}{6}\,\frac{(\kappa/3)\,([\mu_*]_i'-[q_*]_i')}{(\kappa/3)\,([\mu_*]_i-[q_*]_i)}.\label{IM_initc4}
\ea
An evolution free from shell crossings (complying with (\ref{no_xcr})) is obtained with dimensionless initial value functions given by the simple ansatzes  
\bse \label{ICIM}\ba \frac{\kappa[\rho_*]_i}{3h_s^2} &=& 1.2+\frac{0.6}{1+9\,\tan^4 r},\\
  \frac{\kappa [q_*]_i}{3h_s^2} &=& 0.1+\frac{0.05}{1+\tan^2 r}\ea\ese
so that the central and asymptotic values for $\kappa[\mu_*]_i/(3h_s^2)$ are, respectively $1.8$ and $1.2$, while the corresponding values for $\kappa[q_*]_i/(3h_s^2)$ are $0.15$ and $0.1$. Also, we considered the value $w=-0.9$ that is close to a cosmological constant, and $\epsilon=-0.1$, so that CDM particle numbers decay. 

The remaining initial value functions, $[\HH_*]_i,\,\delta_i^{(\rho)},\,\Diq$ and $\Dih$, follow from relations like (\ref{initc2})--(\ref{initc4}), but considering that now $[\KK_*]_i=0$ and $[\mu_*]_i=[\rho_*]_i+[q_*]_i$, and also the coordinate choice (\ref{Yi2}), (\ref{Gdef2}) and (\ref{newdelta2}). Initial conditions  (\ref{ICIM}) yield an evolution free from ``shell
crossings'' (condition (\ref{no_xcr}) is fulfilled).

\end{document}